%% file: H2ff_LMT.tex
\documentclass[11pt]{article}
\pdfoutput=1 

\usepackage{jheppub} 
\usepackage{slashed,bm}
\usepackage{array,multirow,amsmath,latexsym,bm,graphicx}
\allowdisplaybreaks 
\usepackage[normalem]{ulem}
\usepackage[T1]{fontenc} 

\usepackage{mathrsfs}

\def \beq{\begin{equation}}
\def \eeq{\end{equation}}
\def \beqa{\begin{eqnarray}}
\def \eeqa{\end{eqnarray}}

\def \tautau{\tau\bar{\tau}}

\def\msbar{$\overline{\hbox{MS}}$}

\def\sc{{\overline s}}

\def\bea{\begin{eqnarray}}
\def\eea{\end{eqnarray}}

\subheader{\small DCPT/15/34 \\ IPPP/15/17}

\title{One-loop corrections to $h\to b\bar b$ and  
$h\to \tau\bar \tau$ decays in the Standard Model Dimension-6 EFT: four-fermion operators and the large-$m_t$ limit}

\author[a]{Rhorry Gauld,}
\author[a]{Benjamin~D.~Pecjak,}
\author[a]{and Darren~J.~Scott}

\affiliation[a]{Institute for Particle Physics Phenomenology, 
Durham University,  Durham DH1 3LE, UK}

\abstract{ We calculate a set of one-loop corrections to $h\to b\bar
  b$ and $h\to \tau\bar \tau$ decays in the dimension-6 Standard Model
  effective field theory (SMEFT).  In particular, working in the limit
  of vanishing gauge couplings, we calculate directly in the broken
  phase of the theory all large logarithmic corrections and in
  addition the finite corrections in the large-$m_t$ limit.  Moreover,
  we give exact results for one-loop contributions from four-fermion
  operators.  We obtain these corrections within an extension of the
  widely used on-shell renormalisation scheme appropriate for SMEFT
  calculations, and show explicitly how UV divergent bare amplitudes
  from a total of 21 different SMEFT operators are rendered finite
  within this scheme.  As a by-product of the calculation, we also
  compute to one-loop order the logarithmically enhanced and finite
  large-$m_t$ corrections to muon decay in the limit of vanishing
  gauge couplings, which is necessary to implement the $G_F$ input
  parameter scheme within the SMEFT.  }

\emailAdd{rhorry.gauld@durham.ac.uk}
\emailAdd{ben.pecjak@durham.ac.uk}
\emailAdd{d.j.scott@durham.ac.uk}


\begin{document} 
\maketitle
\flushbottom

\section{Introduction}
\label{sec:intro}

One of the main successes of Run-I at the LHC was the
discovery~\cite{Aad:2012tfa,Chatrchyan:2012ufa} of a new particle with
a mass of 125~GeV~\cite{Aad:2015zhl}.  Early measurements of the
various production and decay properties of this particle indicate that
it has quantum numbers ($J^{PC} = 0^{++}$) and coupling strengths to
fermions and gauge bosons consistent with the Standard Model (SM)
Higgs boson~\cite{Bolognesi:2012mm,Aad:2013xqa,Aad:2015gba,Khachatryan:2014jba}.
As the experimental precision of Higgs measurements improves,
comparisons with precise theory calculations will further elucidate
the properties of the observed boson and determine whether they are as
predicted by the SM.

In this paper we study potential new physics contributions to Higgs
boson decays to third generation fermions, namely to $h\to b\bar{b}$
and $h\to \tau\bar{\tau}$ decays.  We perform the analysis within the
framework of the Standard Model Effective Field Theory (SMEFT), where
the effects of new particles at a UV scale $\Lambda_{\rm NP}$ are
parameterised through non-vanishing Wilson coefficients of
higher-dimensional operators.
These operators, which effectively describe the interactions of the
new particles with the SM, are built from gauge invariant combinations
of SM fields and are added onto the usual dimension-4 SM Lagrangian.
The SMEFT approach is justified as
long as the new physics scale $\Lambda_{\rm NP}$ characteristic of the
masses of as yet undiscovered particles is much larger than the
electroweak scale, a scenario which seems quite likely given the
absence of direct evidence for new particles in the Run-I
data.  The main benefit of such an approach is that no assumptions are
made on the nature of new physics, so interpretations of experimental
data can be made in a model-independent fashion.\footnote{When taking
  into account baryon number conserving dimension-6 operators, there
  are $2499$ operators and real parameters~\cite{Alonso:2013hga}.  A
  full global fit of data to such a number of degrees of freedom is
  unrealistic and therefore many simplifying assumptions are made in
  most analyses, but this is a question of implementation rather than
  principle.}

The current precision of Higgs measurements is such that a leading
order (LO) analysis within the SMEFT is sufficient.  However, as the
experimental situation improves (especially at a potential $e^+e^-$
collider, see for example~\cite{Ellis:2015sca}), it will be necessary to carry out next-to-leading order
(NLO) calculations within the SMEFT.  The main point is the
following. When the new physics theory is matched onto the effective
one at the scale $\Lambda_{\rm NP}$, the coefficients of the operators
which are generated in this matching procedure are defined at the
scale $\Lambda_{\rm NP}$.  However, measurements of Higgs couplings
are performed at the scale of the Higgs mass $m_H$ ($m_H \ll
\Lambda_{\rm NP}$).  Under such conditions, renormalisation group (RG)
improved perturbation theory should be used, and the Wilson
coefficients $C_i(\Lambda_{\rm NP})$ should be evolved to the scale
$m_H$ according to the solution to the RG equations, determined from
an anomalous dimension matrix $\gamma_{ij}$.  Since $\gamma_{ij}$ is
in general non-diagonal, the RG evolution (RGE) introduces mixing
among operators. In other words, a measurement of a process which is
sensitive to a particular Wilson coefficient $C_i(m_H)$ in a
LO analysis, is in general sensitive to multiple Wilson
coefficients at the scale $\Lambda_{\rm NP}$, as implied through the
RGE.  In addition, one-loop diagrams also generate non-logarithmic
finite contributions, and there is no way of knowing if these
contributions are large or small without explicitly calculating them.
Both of these effects are neglected in an SMEFT LO analysis, and it is
therefore important to extend analyses to NLO to consistently interpret
experimental data in a robust manner.

From a theoretical point of view, the problem of NLO SMEFT calculations
is interesting in its own right, and there have been several recent theoretical 
advancements in this direction.
In~\cite{Jenkins:2013zja, Jenkins:2013wua,  Alonso:2013hga}, 
the full one-loop anomalous dimension matrix for
baryon number conserving dimension-6 operators was calculated, building
on partial results given 
in \cite{Elias-Miro:2013gya, Elias-Miro:2013mua, Elias-Miro:2013eta}.  The
corresponding analysis for baryon number violating operators was
provided in~\cite{Alonso:2014zka}.  
Such process-independent results form the basis for RG-improved 
LO analyses of physical processes, and are also
integral to the renormalisation procedure used in process-dependent matrix element
calculations such as the one performed in the present work.  Various work 
in such directions can be found in~\cite{Passarino:2012cb,Chen:2013kfa,Grojean:2013kd,
Englert:2014cva,Zhang:2014lya,Pruna:2014asa,Henning:2014wua,Ghezzi:2015vva, David:2015waa, Grober:2015cwa, 
Hartmann:2015oia,Hartmann:2015aia} --- see~\cite{Henning:2014wua,Ghezzi:2015vva} for
detailed discussions on the topic.

In this work, we present results for one-loop corrections to $h\to b
\bar b$ and $h\to \tautau$ decays. The main motivation is the eventual
phenomenological application of the results, however we take the
opportunity to describe in detail how to incorporate the dimension-6
operators into the on-shell renormalisation scheme used in most
Standard Model calculations --- an excellent review of this procedure
is provided in~\cite{Denner:1991kt}.  In order to illustrate this
procedure in the context of the SMEFT, we focus on two types of
contributions. We first calculate the contributions from four-fermion
operators. As this calculation is fairly straightforward, it serves as
a useful example to demonstrate how the renormalisation procedure can
be more generally applied to SMEFT calculations.  After this, we then
compute those contributions which arise in the limit of vanishing
gauge couplings in the broken phase of the theory, where we identify
those terms which are leading in the large-$m_t$ limit.  These limits
are defined more quantitatively below:
\begin{itemize}
\item Vanishing gauge couplings. The QCD corrections, which are
  present for the case of $h\to b\bar b$ decays, are trivially zero.
  For corrections involving electroweak gauge bosons, vanishing gauge
  couplings corresponds to neglecting all contributions which do not
  contain negative powers of $M_{W,Z}^2$, i.e. we calculate terms of
  ${\cal O}(\alpha/M_{W,Z}^2)$.  Consequently, it is not necessary to
  consider real emission diagrams, and the calculation is infrared
  finite.
	
\item Large-$m_t$ limit. To identify the leading-$m_t$ corrections, we
  neglect all fermion masses in the one-loop corrections with the
  exception of the top-quark, and assume $m_t \gg m_H$.  However, as a
  number interesting features of the renormalisation procedure are
  subleading in this limit, we choose to keep the full mass dependence
  in the UV singular contributions and also in the coefficients of
  $\mu$-dependent logarithms.
\end{itemize}
The corrections defined in this way are a well defined subset of the
complete NLO calculation,\footnote{The full results, including the
  dependence on gauge couplings, will be presented in future
  work~\cite{longpaper}.}  and extend the analogous SM
  calculation performed in~\cite{Kniehl:1991ze} to include dimension-6
  contributions.

The layout of the paper is as follows. First, the ingredients of the
SMEFT necessary to compute the tree-level contributions to $h\to b\bar
b$ and $h\to \tautau$ are provided in Section~\ref{sec:tree}.  In
Section~\ref{sec:OL}, we discuss some details of the on-shell
renormalisation scheme in the SMEFT as applied to $h\to f \bar f$
decays, and also comment on how the Fermi constant can be incorporated
as an input parameter.  In Section~\ref{sec:4q}, the contribution from
four-fermion operators is computed.  In Section~\ref{sec:MT}, we
provide the contributions in the limit of vanishing gauge couplings,
applying the large-$m_t$ limit to these corrections.  We discuss 
the phenomenological implications of our results on the interpretation
of future data on $h\to b\bar{b}$ and $h\to \tautau$ decays 
in Section~\ref{sec:Pheno}.  Finally, we give some details of our procedure for
calculating the decay amplitudes in the large $m_t$-limit in
Appendix~\ref{App:LMT}.

\section{Tree-level contributions in the SMEFT}
\label{sec:tree}

In this section we introduce the elements of the SMEFT which are necessary to 
describe the tree-level $h\to b\bar{b}$ and $h\to \tautau$ decay amplitudes. 
We start with the Lagrangian
\begin{align}
\label{eq:Lagrangian}
{\cal L}={\cal L}_{\rm SM} + {\cal L}^{(6)} \, ; \qquad {\cal L}^{(6)} = 
\sum_i C_i(\mu) Q_i(\mu) \,,
\end{align}
which is decomposed into the Standard Model Lagrangian ${\cal L}^{\rm
  SM}$ and dimension-6 Lagrangian ${\cal L}^{(6)}$.  The operators
appearing in the Lagrangian are naturally defined in the unbroken
phase of the gauge theory, where the vacuum expectation value of the
Higgs field vanishes.  A complete set of 59 gauge-invariant
dimension-6 operators was first established in
\cite{Grzadkowski:2010es} (a refinement of the over-complete basis
originally proposed in~\cite{Buchmuller:1985jz}), and is listed in
Table~\ref{op59}.  The Wilson coefficients $C_i$ of the dimension-6
operators implicitly contain two inverse powers of $\Lambda_{\rm NP}$,
and are therefore dimensionful. Additionally, the labeling convention
of the operators appearing in Table~\ref{op59} is also applied to the
corresponding Wilson coefficient. For example, the Wilson coefficient
of the operator $Q_{dH}$ is $C_{dH}$.  This notation will be used
throughout.

\subsection{Yukawa sector}
The effective Yukawa couplings and mass matrices in the broken phase
of the theory, where the vacuum expectation value of the Higgs field 
is non-vanishing, arise from the following terms in the unbroken one: 
\begin{align}
\mathcal{L} =& -\left[ [Y_u]_{rs} \tilde{H}^{\dagger j} \overline u_r \,   Q_{sj} + [Y_d]_{rs} H^{\dagger j} \overline d_r \,   Q_{sj} + [Y_e]_{rs} H^{\dagger j} \overline e_r \,   L_{sj}+ h.c. \right] 
\nonumber \\[1mm] &
+\left[ C^*_{\substack{u H \\ sr}} (H^\dagger H)  \tilde{H}^{\dagger j} \overline u_r \,   Q_{sj} + C^*_{\substack{d H \\ sr}} (H^\dagger H) H^{\dagger j} \overline d_r \, Q_{sj} + C^*_{\substack{e H \\ sr}} (H^\dagger H) H^{\dagger j} \overline e_r \, L_{sj} +h.c.\right]
\nonumber \\[1mm] &
-V(H)  \, ,
\end{align}
where
\begin{align}  \label{Hpot}
V(H) = \lambda \left(H^\dagger H-\frac12 v^2\right)^2 - C_H (H^\dagger H)^3 \,.
\end{align}
The dimension-6 operators alter the tree level-relations between
parameters in the broken and unbroken phase of the theory compared to
the SM.  We now summarise the modifications relevant for $h\to b\bar b$ and 
$h\to \tautau$ decay amplitudes, following
closely the discussion and notation from~\cite{Alonso:2013hga}, which
contains all necessary elements.  

We write the Higgs doublet in a general $R_\xi$ gauge in the
broken phase of the theory as
\begin{align} \label{eq:Higgs}
 H(x) = \frac{1}{\sqrt{2}}\left( \begin{array}{c}
-\sqrt{2}i\phi^+(x) \\
 \left[1+C_{H,{\rm kin}}\right]h(x) 
+i\left[1-\frac{v^2}{4}C_{HD}\right] \phi^0(x) + v_T\end{array}  \right),
 \end{align}
 where $\phi^0$ and $\phi^{+}$ are Goldstone boson modes, and the
 following relations have been introduced
\begin{align} \label{eq:Class3}
C_{H,{\rm kin}} \equiv \left(C_{H\Box}-\frac{1}{4}C_{HD}\right)v^2 \,, \qquad 
v_T \equiv \left(1+\frac{3 C_H v^2}{8\lambda}\right)v \,.
\end{align}
The prefactors of the $h(x)$ and $\phi^{0}(x)$ fields are determined
by the requirement that the kinetic terms in the broken phase of the
theory are canonically normalised. We have distinguished the quantities
$v$ and $v_T$ above, but since the difference between them is a dimension-6
effect, they can be interchanged freely when multiplying a dimension-6
Wilson coefficient, and we will always refer to this quantity as $v_T$ under
such circumstances.

The Higgs boson mass is found by expanding (\ref{Hpot}), and leads to
\begin{align} \label{eq:Hmass}
m_H^2 = 2 \lambda v_T^2\left(1- \frac{3 C_H v^2}{2\lambda}+2 C_{H,\rm{kin}}\right) \, .
\end{align}
Similarly, the effective mass and Yukawa matrices for fermions are 
\begin{align} \label{eq:Yuk}
\left[M_f\right]_{rs}& = \frac{v_T}{\sqrt{2}}
 \left( [Y_f]_{rs} - \frac{1}{2}v_T^2 C^*_{\substack{f H \\ sr}}  \right), \\
\left[{\cal Y}_f\right]_{rs}& =
 \frac{1}{\sqrt 2}\left( [Y_f]_{rs}\left[1+C_{H,{\rm kin}}\right] 
- \frac{3}{2}v_T^2 C^*_{\substack{f H \\ sr}} \right) \nonumber \\
&= \frac{1}{v_T}\left[M_f\right]_{rs}\left[1+C_{H,{\rm kin}}\right]
-\frac{v_T^2}{\sqrt{2}} C^*_{\substack{f H \\ sr}} \,.
\end{align}
The Yukawa and mass matrices depend on distinct linear combinations of the
SM Yukawa matrix and the dimension-6 terms $C^*_{fH}$.
Therefore, after transforming from the gauge to mass eigenstates by
performing field redefinitions on the fermion fields, the operators in
the mass basis contain a myriad of flavour violating effects beyond
those in the CKM matrix.  While such flavour violating effects 
beyond those present in the SM are interesting
phenomenologically (see for example~\cite{Harnik:2012pb}),
particularly in light of the excess observed in $h\to \tau
\mu$ events by the CMS~\cite{Khachatryan:2015kon} collaboration, the
main focus of the present work is on one-loop corrections rather than
questions of flavour.  Therefore, we ignore such flavour-violating couplings
in this work.  This can be made more rigorous by imposing minimal
flavour violation (MFV)~\cite{Chivukula:1987py,D'Ambrosio:2002ex}, an
assumption which ensures that the mass and Yukawa matrices are
simultaneously diagonalizable at all scales, a feature preserved by
the RG running~\cite{Alonso:2013hga}. The transition from the
gauge to mass eigenstates then proceeds much as in the SM, and in
fact can be rendered trivial by considering only the third generation
in the calculation of one-loop effects.  We will use this set up
throughout the paper, i.e.  consider one generation of fermions and
set the CKM element $V_{tb}$ to unity.

With these simplifications in place, the Yukawa couplings 
to third generation fermions, defined as the coefficients 
of the $hf\bar f$ coupling in the mass basis of the broken theory, 
are  related to the physical masses according to 
\begin{align}\label{eq:YMrel}
y_{f} = \sqrt{2}\frac{m_{f}}{v_T} + \frac{v_T^2}{2} C^*_{fH} \,,
\end{align} 
and it is a simple matter to calculate the tree-level decay amplitude
for the process $h\to f\bar{f}$:
\begin{align}
i{\cal M}^{(0)}(h\to f\bar f)= 
-i \bar u(p_f) 
\left({\cal M}_{f,L}^{(0)} P_{L} + {\cal M}_{f,L}^{(0)*} P_R\right)v(p_{\bar f}) \, ,
\end{align}
where
\begin{align}
\label{eq:Mtree}
{\cal M}_{f,L}^{(0)} &=  
\frac{m_f}{v_T}\left[ 1+C_{H,{\rm kin}} \right]-  \frac{ v_T^2}{\sqrt{2}}C_{fH}^* \, .
\end{align}

\subsection{Input parameters}
\label{sec:Inputs}
We have expressed the result (\ref{eq:Mtree}) in terms of $v_T$, but
in practice this parameter should be eliminated in terms of
observables of the broken phase of the theory.  In the renormalisation
procedure discussed in the next section, we choose to work with the
following set of independent, physical parameters: 
\begin{align}
 \label{param}
\bar{e}, m_H, M_W, M_Z, m_f, C_i\, . 
\end{align} 
Using the expressions from~\cite{Alonso:2013hga}, one has
\begin{align}
M_W^2 & = \frac{\bar{g}_2^2 v_T^2}{4} \, , \nonumber \\
M_Z^2 & = \frac{v_T^2}{4}(\bar{g}_1^2 + \bar{g}_2^2)
+\frac{1}{8} v_T^4 C_{HD}(\bar{g}_1^2+\bar{g}_2^2)
+\frac{1}{2} v_T^4 \bar{g}_1 \bar{g}_2 C_{HWB}\,, \nonumber \\
\bar{e}&= \bar{g}_2 \bar{s}_w -\frac{1}{2}\bar{c}_w \bar{g}_2v_T^2 C_{HWB} \, , 
\nonumber \\
\bar{s}_w^2 &= \frac{\bar{g}_1^2}{\bar{g}_1^2+\bar{g}_2^2}
+ \frac{\bar{g}_1\bar{g}_2(\bar{g}_2^2-\bar{g}_1^2)}{(\bar{g}_1^2+\bar{g}_2^2)^2}
v_T^2 C_{HWB} \, .
\end{align}
The barred quantities appear as couplings in the covariant derivative
in the broken phase of the theory after rotation to the mass
eigenbasis; in particular $\bar{g}_2$ governs the charged current
couplings while $\bar{e}$ is the electric charge.  Manipulating the
above expressions we can write
\begin{align}
\label{eq:vhat_def}
\frac{1}{v_T} = \frac{\bar{e}}{2M_W \bar{s}_w}\left(1+
\frac{\hat{c}_w}{2\hat{s}_w} C_{HWB} \hat{v}_T^2\right) \, ,
\end{align} 
where we have defined
\begin{align}
\label{eq:OnShellReps}
\hat{v}_T \equiv \frac{2 M_W \hat{s}_w}{\bar e} \, ;
\qquad \hat{s}_w^2 \equiv 1-\frac{M_W^2}{M_Z^2} \,, \quad  \hat{c}_w^2 \equiv 1-\hat{s}_w^2 \,,
\end{align}
such that the hatted quantities are the usual definitions in the SM.
In expression~(\ref{eq:vhat_def}), we denote parameters multiplying
the Wilson coefficients by the hatted quantities. This is consistent
to $\mathcal{O}(1/\Lambda_{NP}^2)$, and is the notation which will be
adopted throughout this work.  It is possible to re-express $v_T$
and $\bar{s}_w$ in terms of the gauge boson masses, and quantities
derived from them. In particular, the quantity $\bar{s}_w$ can be
expressed as
\begin{align}
\label{eq:sbar_def}
\bar{s}^2_w = \hat{s}^2_w - \frac{\hat{c}_w^2 v_T^2}{2} \left(C_{HD}+ 2 \frac{\hat{s}_w}{\hat{c}_w} \, C_{HWB}\right) \,,
\end{align}
and inserting this into (\ref{eq:vhat_def}) leads to
\begin{align}
\label{eq:vhat}
\frac{1}{v_T} = \frac{1}{\hat{v}_T}
+\frac{\hat{c}_W}{\hat{s}_W}
\left(C_{HWB}+\frac{\hat{c}_W}{4\hat{s}_W} C_{HD}  \right)\hat{v}_T \,.
\end{align}
Equation~(\ref{eq:vhat}) then allows to write $v_T$ in terms of the
parameters (\ref{param}), that is, the physical parameters in the
broken phase of the theory.  While this is a reasonable choice, it is
instead customary to eliminate $M_W$ in favour of the Fermi constant
$G_F$, defined and extracted through the muon decay
rate~\cite{Agashe:2014kda}.  At tree level, and ignoring contributions
which do not interfere with the SM,  we can write
\cite{Alonso:2013hga}
\begin{align} \label{eq:GF}
 \frac{1}{\sqrt{2}} \frac{1}{v_T^2}= G_F  
-\frac{1}{\sqrt{2}}\left(C_{\substack{Hl \\ ee}}^{(3)}+C_{\substack{Hl \\ \mu\mu}}^{(3)} \right) + \frac{1}{2\sqrt{2}}\left(C_{\substack{ll \\ \mu ee \mu}} 
+ C_{\substack{ll \\ e\mu \mu e}} \right) \, ,
\end{align}
where the operator $C_{\substack{Hl \\ ll}}^{(3)}$, which alters the
$W$ boson coupling to the lepton doublets, and also the four-fermion
operators $C_{\substack{ll \\ e\mu \mu e}}$ and $C_{\substack{ll \\
    \mu ee \mu}}$ explicitly enter the amplitude for muon decay.  One
can then insert the above equation into (\ref{eq:vhat}) and solve for
$M_W$ as a function of $G_F$ and the other observables appearing in
(\ref{param}).
 
\section{The one-loop renormalisation procedure}
\label{sec:OL}

From a practical point of view, the calculation of one-loop
corrections to $h\to b\bar b$ and $h\to \tautau$
decays in the SMEFT has two components --- the bare one-loop matrix
elements, and the UV counterterms required to subtract the UV poles
(and in some cases finite parts) from these divergent matrix elements. The
calculation of the one-loop matrix elements is conceptually
straightforward and will be discussed later on.  In this section we
cover the somewhat more subtle issue of constructing the UV
counterterms.  In particular, we explain how to adapt the on-shell
renormalisation scheme used to calculate electroweak corrections in
the SM to the SMEFT case.

To renormalise bare amplitudes we must provide UV counterterms for the
set of independent, physical parameters in (\ref{param}), and also
perform wavefunction renormalisation on external fields.  We choose to
renormalise the masses and electric charge in the on-shell scheme, and
construct counterterms related to these quantities exactly as in the
SM.  This requires the computation of a number of two-point functions
directly in the broken phase of the theory.  On the other hand, we
renormalise the Wilson coefficients $C_i$ in the \msbar~scheme, as is
standard in EFT calculations. Crucially, the counterterms associated
with the Wilson coefficients can be taken from results in the unbroken
phase of the theory calculated
in~\cite{Jenkins:2013zja,Jenkins:2013wua,Alonso:2013hga}.

We begin with wavefunction, mass, and electric charge renormalisation,
which proceeds as in the SM. We will only discuss those contributions
relevant for $h\to b\bar{b}$ and $h\to \tautau$ decays. Defining the
renormalised fields in terms of bare ones, indicated with the
superscript $(0)$, we have 
\begin{align}
h^{(0)} &=\sqrt{Z_h} h = \left(1+\frac12 \delta Z_h\right)h \,, \nonumber \\
f^{(0)}_L & = \sqrt{Z^{L}_f} f_L = \left(1+\frac12 \delta Z^L_f\right)f_L \,, \nonumber \\
f^{(0)}_R &= \sqrt{Z^{R}_f} f_R = \left(1+\frac12 \delta Z^R_f\right)f_R \,,
\end{align}
where the fermion subscript $f$ refers to either 
$b$-quarks or $\tau$-leptons. We define renormalised 
masses and  the renormalised electric charge as 
\begin{align}
M^{(0)} = M + \delta M, \qquad \bar{e}_0 = \bar{e} + \delta \bar{e} \, ,
\end{align}
where $M$ is a generic mass.
The Higgs mass does not enter our tree-level expression for the decay 
rate, and it is therefore not renormalised. As a consequence, the contribution
from tadpole diagrams are cancelled exactly by those of the corresponding 
counterterms, and these contributions can therefore be effectively ignored in our calculation.

To determine wave function renormalisation factors
and the counterterms related to mass and electric charge
renormalisation, we follow the procedure outlined 
in~\cite{Denner:1991kt,Kniehl:1996bd}, which requires
the computation of a set of two-point functions in the broken
phase of the theory.  
We write the generic two-point functions as
\begin{align}
\Gamma^f(p) &= i (\slashed{p} - m_f)+
i \left[\slashed{p}\left( P_L \Sigma_f^L(p^2) + P_R\Sigma_f^R(p^2)\right)
+ m_f\left( \Sigma_f^S(p^2)P_L +  \Sigma_f^{S*}(p^2)P_R \right)\right] \, ,
\nonumber \\ 
\Gamma^H(k) &= i(k^2-m_H^2)+ i \Sigma^H(k^2)  \, ,
\nonumber \\
\Gamma^W_{\mu\nu}(k) &= -ig_{\mu\nu}(k^2-M_W^2) 
-  i\left(g_{\mu\nu}-\frac{k_\mu k_\nu}{k^2} \right) \Sigma^W_T(k^2)
- i \frac{k_\mu k_\nu}{k^2} \Sigma^W_L(k^2) \,,
\nonumber \\
\Gamma^{ab}_{\mu\nu}(k) &= -ig_{\mu\nu}(k^2-M_a^2) \delta_{ab}
-  i\left(g_{\mu\nu}-\frac{k_\mu k_\nu}{k^2} \right) \Sigma^{ab}_T(k^2)
- i \frac{k_\mu k_\nu}{k^2} \Sigma^{ab}_L(k^2) \,,
\end{align}
where $a,b$ = $A,Z$, and $M_A^2 = 0$.
Given results for the two-point functions, one can 
calculate the counterterms from wavefunction renormalisation
in the on-shell scheme according to\footnote{We follow the
 convention of \cite{Kniehl:1996bd} and choose $\delta Z_f^{R}$ to be real.}

\begin{align}
\label{eq:delZ}
\delta Z_f^L = &-\widetilde{\rm Re}\, \Sigma_f^L(m_f^2)+ \Sigma_f^S(m_f^2)-\Sigma_f^{S*}(m_f^2) \nonumber  \\
& -m_f^2 \frac{\partial}{\partial p^2}\widetilde{\rm Re}
\left[\Sigma_f^{L}(p^2)+ \Sigma_f^{R}(p^2)
+ \Sigma_f^{S}(p^2) + \Sigma_f^{S*}(p^2) \right] \bigg |_{p^2=m_f^2}  \,,  \nonumber \\
\delta Z_f^{R}& =  -\widetilde{\rm Re}\, \Sigma^{f,R}(m_f^2)  \nonumber \\
&  -m_f^2 \frac{\partial}{\partial p^2}\widetilde{\rm Re}
\left[\Sigma_f^{L}(p^2)+ \Sigma_f^{R}(p^2)
+ \Sigma_f^{S}(p^2) + \Sigma_f^{S*}(p^2) \right] \bigg |_{p^2=m_f^2}\,, \nonumber \\
\delta Z_h & = -{\rm Re}\frac{\partial \Sigma^H(k^2)}{\partial k^2}\bigg|_{k^2=m_H^2} \, .
\end{align}
The mass counterterms are computed as
\begin{align}
\label{eq:delm}
\delta m_f &= \frac{m_f}{2} \widetilde{\rm Re}\left(\Sigma_f^{L}(m_f^2)+ \Sigma_f^{R}(m_f^2)
+ \Sigma_f^{S}(m_f^2)+ \Sigma_f^{S*}(m_f^2)\right) \, , \nonumber \\
\frac{\delta M_W}{M_W} &=  \widetilde{\rm Re}\, \frac{\Sigma^W_T(M_W^2)}{2M_W^2} \,.
\end{align}
We have listed relations for the $W$ boson above -- those for the $Z$ boson 
are completely analogous.
The symbol $\widetilde{\rm Re}$ takes the real part of the matrix
elements in the two-point functions but not of the CKM matrix elements
or the Wilson coefficients themselves.
The renormalisation of the electric charge is also computed from two-point functions
according to
\beq \label{eq:Charge} \frac{\delta \bar{e}}{\bar{e}} = \frac{1}{2}
\frac{\partial \Sigma_T^{AA}(k^2)}{\partial k^2} \bigg|_{k^2=0} +
\frac{(v_f - a_f)}{Q_f}\frac{\Sigma_T^{AZ}(0)}{M_Z^2} \, , \eeq where
$v_f$ and $a_f$ are the vector and axial coupling of the $Z$ boson to
fermions and $Q_f$ is the fermion electric charge. 
In the SM the difference between the vector and axial couplings is $v_f-a_f
= - Q_f s_w/c_w$, which when inserted into~(\ref{eq:Charge}) leads to
the usual relation for electric charge renormalisation~\cite{Denner:1991kt}. 
In the SMEFT, the expression for $v_f-a_f$ is altered by dimension-6
contributions. However, the quantity $\Sigma_T^{AZ}(0)$ itself is
subleading in the limit of vanishing gauge couplings and so the exact
form of $v_f-a_f$ is irrelevant to what follows.

We next turn to counterterms related to operator renormalisation.  At
the level of the Lagrangian, such counterterms have the form $\delta
C_i Q_i$, where $\delta C_i = \sum_j \gamma_{ij} C_j$ and thus
involves a linear combination of all Wilson coefficients in the basis.
We need such counterterms for each operator appearing in the
tree-level expression (\ref{eq:Mtree}). To one-loop order in the
\msbar~scheme, we can write
\begin{align}
C_i^{(0)} = C_i(\mu) +\frac{ \delta C_i(\mu)}{16\pi^2} = C_i(\mu)+\frac{1}{2\hat{\epsilon}}\frac{1}{16\pi^2} \dot{C}_i(\mu)
 \,,
\end{align}
where we have defined
\begin{align}
\label{eq:delCdef}
\dot{C}_i(\mu) \equiv 16 \pi^2 \left(\mu \frac{d}{d\mu}C_i(\mu) \right)  \,.
\end{align}
It is understood that we evaluate the right-hand side of the above
equation at one-loop order using the results from
\cite{Jenkins:2013zja, Jenkins:2013wua, Alonso:2013hga}.
We have also introduced the notation
\begin{align}
\label{eq:epshat}
\frac{1}{\hat{\epsilon}} \equiv \frac{1}{\epsilon}- \gamma_E + \ln(4\pi) \, ,
\end{align}
where $\epsilon$ is the dimensional regulator for integrals evaluated
in $d=4-2\epsilon$ dimensions. UV divergences in loop integrals always
appear as factors of $1/\hat{\epsilon}$, and rather than clutter
notation we shall  write such factors as $1/\epsilon$ in the
rest of the paper, with the understanding that such poles are
accompanied by the universal, finite terms on the right-hand side of
(\ref{eq:epshat}). When the UV poles of the bare and counterterm
matrix elements are cancelled, so too are these constant terms.

With these ingredients in place, we can now construct the explicit
form of the UV counterterms for the specific case of $h\to f\bar f$.
We  take the tree-level expression (\ref{eq:Mtree}), interpret
the quantities in it as bare parameters, and then replace these bare
parameters by the renormalised ones.  For the vacuum expectation
value $v_T$, this leads us to write
\begin{align}
\frac{1}{v_T^{(0)}} = \frac{1}{v_T}\left(1-\frac{\delta v_T}{v_T} \right)  \, .
\end{align}
We can derive an explicit expression for $\delta v_T$ as a function of
the counterterms for the physical observables
(\ref{param}) using (\ref{eq:vhat}).  Defining
\begin{align}
\frac{\delta \hat{c}_w}{\hat{c}_w}&\equiv \frac{\delta M_W}{M_W}-\frac{\delta M_Z}{M_Z} ,
\, \ \quad
\frac{\delta \hat{s}_w}{\hat{s}_w} \equiv -\frac{\hat{c}^2_w}{\hat{s}_w^2}\frac{\delta \hat{c}_w}{\hat{c}_w} \, ,
\nonumber \\ 
\frac{\delta \hat{v}_T}{\hat{v}_T}&\equiv \frac{\delta M_W}{M_W} + \frac{\delta \hat{s}_w}{\hat{s}_w} - \frac{\delta \bar{e}}{\bar{e}}\,.
\end{align}
we find that
\begin{align}
\frac{\delta v_T}{v_T} &= \frac{\delta M_W}{M_W} 
+ \frac{\delta \bar{s}_w}{\bar{s}_w} - \frac{\delta \bar{e}}{\bar{e}} 
- \frac{\hat{v}_T^2\hat{c}_w }{2\hat{s}_w} \delta C_{HWB}
\nonumber \\ &
-  \frac{\hat{c}_w}{2\hat{s}_w}\hat{v}_T^2
\left(\frac{\delta \hat{c}_w}{\hat{c}_w} -\frac{\delta \hat{s}_w}{\hat{s}_w}
+2\frac{\delta \hat{v}_T}{\hat{v}_T}\right) C_{HWB} \,.
\end{align}
The counterterm for $\delta\bar{s}_w$ can be computed using
(\ref{eq:sbar_def}). One has
\begin{align}
\label{eq:dsbar}
\frac{\delta\bar{s}_w}{\bar{s}_w}&= \frac{\delta \hat{s}_w}{\hat{s}_w}
-\frac{\hat{c}_w^2}{2\hat{s}_w^2} \hat{v}_T^2\left(\frac{\delta \hat{c}_w}{\hat{c}_w} - \frac{\delta \hat{s}_w}{\hat{s}_w} +\frac{\delta \hat{v}_T}{\hat{v}_T}  \right)C_{HD} - \frac{\hat{c}_w^2 \hat{v}_T^2}{4 \hat{s}_w^2} 
\delta C_{HD} \nonumber \\ 
&- \frac{\hat{v}_T^2\hat{c}_w }{2\hat{s}_w} \delta C_{HWB}
-  \frac{\hat{c}_w}{2\hat{s}_w}\hat{v}_T^2
\left(\frac{\delta \hat{c}_w}{\hat{c}_w} -\frac{\delta \hat{s}_w}{\hat{s}_w}
+2\frac{\delta \hat{v}_T}{\hat{v}_T}\right) C_{HWB} \,.
\end{align}

Note that the two-point functions, and the renormalisation
counterterms derived from them as discussed above
receive both SM and dimension-6 contributions. We make this explicit
by defining expansion coefficients according to
\begin{align}
\delta Z &= \frac{1}{16\pi^2}\left( \delta Z^{(4)} + \delta Z^{(6)}\right) +\dots \,,
\end{align}
and similarly for $\delta M$, $\delta \bar{e}$ and $\delta v_{T}$.
The superscript $(4)$ then refers to SM contributions, while the
superscript $(6)$ refers to dimension-6 contributions.  The counterterms
$\delta C_i$ related to operator renormalisation are purely dimension-6,
so we do not label them with a (redundant) superscript $(6)$.    

The counterterm for the $h\to f\bar{f}$ decay
amplitude can now be written as 
\begin{align} \label{CT}
i{\cal M}^{\rm C.T.}(h\to f\bar f)= 
-i\bar u(p_f) 
\left(\delta {\cal M}_L P_L + \delta {\cal M}_L^*  P_R\right) v(p_{\bar f}) \,,
\end{align}
where we distinguish SM and dimension-6 contributions through the notation 
\begin{align}
\delta {\cal M}_L &= \frac{1}{16\pi^2}\left(\delta {\cal M}_L^{(4)} + \delta {\cal M}_L^{(6)}\right) 
+ \dots \, .
\end{align}
The SM contributions read
\begin{align}
\label{eq:dim4CT}
\delta {\cal M}_L^{(4)} & = \frac{m_f}{v_T}
\left(\frac{\delta m_f^{(4)}}{m_f}-\frac{\delta v_T^{(4)}}{v_T}+\frac{1}{2} \delta Z_h^{(4)} +\frac{1}{2}\delta Z_f^{(4),L} + \frac{1}{2}\delta Z_f^{(4),R*} \right) \, ,
\end{align}
and the dimension-6 contributions are 
\begin{align}
\label{eq:dim6CT}
\delta {\cal M}_L^{(6)} =& 
\left( \frac{m_f}{v_T} C_{H,{\rm kin}} \right)  \left(\frac{\delta m_f^{(4)}}{m_f} 
-\frac{\delta v_T^{(4)}}{v_T}+\frac12 \delta Z_h^{(4)}+ \frac12 \delta Z_f^{(4),L}+ \frac{1}{2} \delta Z_f^{(4),R*}\right) \nonumber \\ 
&- \frac{v_T^2}{\sqrt{2}} C_{bH}^*  \left(2 \frac{\delta v_T^{(4)}}{v_T}+\frac12 \delta Z_h^{(4)}+ \frac12 \delta Z_f^{(4),L}+ \frac{1}{2} \delta Z_f^{(4),R*}\right) \nonumber \\
&+\frac{m_f}{v_T}\left(\frac{\delta m_f^{(6)}}{m_f} 
-\frac{\delta v_T^{(6)}}{v_T}+\frac12 \delta Z_h^{(6)}+ \frac12 \delta Z_f^{(6),L}+ \frac{1}{2} \delta Z_f^{(6),R*}\right) \nonumber \\ 
& +\frac{m_f}{v_T} \delta C_{H,{\rm kin}} - \frac{v_T^2}{\sqrt{2}} \delta C_{fH}^* \,.
\end{align}
Clearly, for the $h\to b\bar b$ matrix element it is necessary to
include the $b$-quark mass ($\delta m_b/m_b$) and wavefunction
($\delta Z_b$) renormalisation factors in the counterterm, while for
the $h\to \tautau$ matrix element the corresponding $\tau$-lepton
factors should be included. The above results~(\ref{eq:dim6CT}) are
valid in the \msbar~scheme for the Wilson coefficients, and the
on-shell scheme (pole scheme) for the masses and the electric charge.
One may instead wish to use different definitions for these masses,
such as the \msbar~scheme, which shuffles finite contributions between the
matrix elements and the masses. We will provide an example on how this
can be done when we consider four-fermion contributions in 
Section~\ref{sec:4q}.

The procedure to calculate the one-loop corrections to the 
$h\to f\bar f$ decay rate in a given renormalisation scheme is now 
clear. Compute
\begin{align}
\label{eq:hbbren}
{\cal M}^{(1)}(h\to f\bar f) =  {\cal M}^{(1),{\rm bare}} +{\cal M}^{\rm C.T.} \,,
\end{align}
where each of the terms receives both SM and dimension-6
contributions.  This procedure is straightforward to implement in the
case where the parameters (\ref{param}) are used as input.  However,
as mentioned in Section~\ref{sec:Inputs}, it is customary to eliminate
$M_W$ dependence in favour of the Fermi constant $G_F$ as measured from
muon decay.  In order to do so we must modify the tree-level
relation~(\ref{eq:GF}) to a form appropriate at one-loop.  We do this
by writing
\begin{align} \label{eq:GF_one_loop}
\frac{1}{\sqrt{2}} \frac{1}{v_T^2}\left(1+\Delta r \right)  = G_F
+ \Delta R^{(6)}   \, .
\end{align}
The expression for $\Delta r$, which summarises the finite non-QED
radiative corrections to muon decay in terms of two-point functions
can be found in~\cite{Sirlin:1980nh}.  The contributions labelled as
$\Delta R^{(6)}$ summarise the finite process specific contributions
to muon decay in the SMEFT.  Evaluating the expression for $\Delta r$
in the limit of vanishing gauge-couplings, we find that
\begin{align}
\label{eq:Deltar}
\Delta r =   2\left(\frac{\delta M_W}{M_W} - \frac{\delta v_T}{v_T} \right)\,.
\end{align}
To implement this scheme to one-loop order, we first define expansion
coefficients as
\begin{align}
\label{eq:Rcoeffs}
\Delta r & = \frac{1}{16\pi^2}\left(\Delta r^{(4,1)}+ \Delta r^{(6,1)} \right)\, , \nonumber \\
\Delta R^{(6)} & = \Delta R^{(6,0)} + \frac{1}{16\pi^2}  \Delta R^{(6,1)} \, .
\end{align}
The tree-level piece $\Delta R^{(6,0)}$ is obtained by matching with
(\ref{eq:GF}).  We shall give explicit expressions for the one-loop
corrections to $\Delta r$ and $\Delta R^{(6)}$ in
Section~\ref{sec:MT}, where we also give explicit results for the
renormalised one-loop decay amplitude after eliminating $v_T$
dependence using (\ref{eq:GF_one_loop}).  For now, we simply note that  the
counterterms derived after writing $v_T$ in terms of $G_F$ take a very
simple form.  They can be obtained from (\ref{eq:dim4CT}) and
(\ref{eq:dim6CT}) by replacing $\delta v_T/v_T$ with $\delta M_W/M_W$,
which follows from the definition of $\Delta r$, and then in addition
adding on the extra dimension-6 pieces contained in $\Delta R^{(6)}$
by hand.

\section{The one-loop contribution from four-fermion operators}
\label{sec:4q}

In this section we compute the one-loop contributions from
four-fermion operators to $h\to b\bar{b}$ and $h\to \tautau$
decays. Not only are these the simplest dimension-6 contributions
to compute, we will see in Section~\ref{sec:Pheno} that they are among
the most important numerically. At the same time, their calculation
nicely illustrates many aspects of the renormalisation procedure
outlined in the previous section.  

The list of operators which must be considered are those labelled as
Class `8' in Table~\ref{op59}.  In general, the coefficients of the
four-fermion operators carry four flavour indices labeling the fermion
generations.  In the current study, we consider only $b$-quark and
$\tau$-lepton final states, and only the radiative corrections due the
third generation field content are considered, and consequently these
flavour indices are redundant and will be dropped in what follows.
For example, the scalar operator $(\bar{L}R)(\bar R L)$ is labelled
as $Q_{l\tau bq} = (\bar{l}^j \tau)(\bar{b}q_j)$.

It is convenient to calculate the one-loop corrections by performing
Passarino-Veltmann reduction~\cite{Passarino:1978jh} and writing 
the results in terms of the standard one-loop scalar integrals.
In order to make explicit the UV divergent parts of these integrals, we
write the one-loop scalar integrals as
\begin{align}
\label{eq:ABints}
A_0\left(s\right)&=\frac{s}{\epsilon} + \hat{A}_0(s) \, ,\\
B_0\left(s,m_1^2,m_2^2\right) &= \frac{1}{\epsilon} + \hat{B}_0(s,m_1^2,m_2^2) \,,
\end{align}
where we have defined the finite, $\mu$-dependent integrals
\begin{align} 
\hat{A}_0(s) & = s - s\ln\left(\frac{s-i0}{\mu^2}\right)\, , \\
\hat{B}_0(s,m_1^2,m_2^2) & =
2 -\log \left(\frac{s-i0}{\mu^2}\right)+\sum_{i=1}^2 \left[\lambda_i \ln\left(\frac{\lambda_i-1}{\lambda_i}\right)-\ln\left(\lambda_i-1\right)\right] \, , \label{eq:Bint}
\end{align}
and
\begin{align}
\lambda_i = \frac{s-m_2^2 + m_1^2 \pm\sqrt{(s-m_2^2+m_1^2)^2 -4s(m_1^2-i0)}}{2s} \,.
\end{align}
In section~\ref{sec:MT}, when we consider the large-$m_t$ limit, we
will also use explicit results for special values of the arguments of
the triangle integral $C_0$.  These results can be obtained
from~\cite{Kniehl:1991ze}, and are provided in
Appendix~\ref{App:LMT}.

\subsection{Bare matrix element}
\label{sec:4qME}
\begin{figure}[t]
\centering
\includegraphics[scale=1.3]{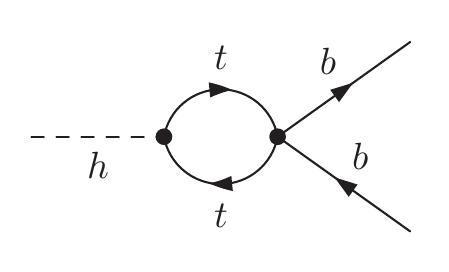}
\hspace{1cm}
\includegraphics[scale=1.3]{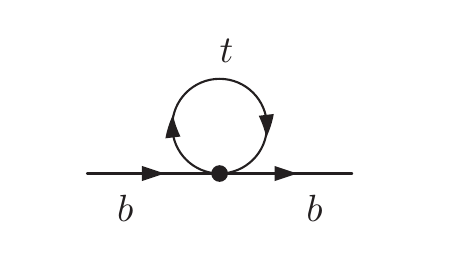}
\caption{Examples of one-loop diagrams involving Class 8 operators to
  the $h\to b\bar b$ process (left), and to the $b$-quark
  two-point function (right).  The corresponding diagrams for
  $h\to\tautau$ are of similar form.}
\label{fig:4q}
\end{figure}

We begin by computing the contribution from the four-fermion operators
to the bare matrix element. The four-fermion operators do not
contribute to the tree-level result ~(\ref{eq:Mtree}), and so it is
only necessary to evaluate the one-loop contributions.  The relevant
diagrams are of the form of that shown in the left-hand side of
Figure~\ref{fig:4q}.  The contributions from the vector operators
$(\bar{L}L)(\bar L L)$ and $(\bar{R}R)(\bar R R)$ vanish due to their
Dirac structure.  We write the non-vanishing contribution for the sum
of all four-fermion diagrams to the bare matrix element as
\begin{align}
i{\cal M}^{(1),{\rm bare}}_{8}(h\to f\bar f)= 
- i \frac{1}{16\pi^2} \bar u(p_f) 
\left( C^{L,(1),{\rm bare}}_{8,f} P_L + C^{R,(1),\rm bare}_{8,f} P_R\right) v(p_{\bar f}) \, .
\end{align}

\noindent
For $h\to b\bar b$ decays one finds
\begin{align}
C^{L,(1),\rm bare}_{8,b} &=
\frac{1}{v_T} \Bigg[
m_b (4-2 \epsilon) 
I_8^b   \left(C_{qb}^{(1)}+c_{F,3} C_{qb}^{(8)}\right)
- 2 m_{\tau}
I_8^{\tau}  C_{l\tau bq}
\nonumber \\ &
- m_t 
I_8^t  \left( (2 N_c+1) C^{(1)*}_{qtqb} + c_{F,3} C^{(8)*}_{qtqb}\right)
\Bigg] \, ,
\end{align}
and for $h\to \tautau$ one has 
\begin{align}
C^{L,(1),\rm bare}_{8,\tau} &=
\frac{1}{v_T} \Bigg[
m_{\tau} (4-2 \epsilon) I_8^{\tau} C_{l\tau}
+ 2 N_c m_t  I_8^{t} C^{(1)*}_{l\tau qt}
- 2 N_c m_b I_8^b  C^*_{l\tau bq}
\Bigg] \,.
\end{align}
Results for the functions $C_{8,f}^{R,(1),\rm bare}$ are obtained through 
the relation
\begin{align}
\label{eq:CRs}
C_{8,f}^{R,(1),\rm bare} = C_{8,f}^{L,(1),\rm bare}\left( C_i^*\leftrightarrow  C_i\right)\,,
\end{align}
which clearly only effects the complex Wilson coefficients, i.e. those
multiplying $(\bar{L}R)(\bar{R} L)$ or $(\bar{L}R)(\bar{L} R)$
operators which are labeled with four subscripts.  In the above expressions for
the bare matrix elements, the following notation has been introduced
\begin{align}
I^j_{8} &=  A_0(m_j^2)-\frac{1}{2}\left(m_H^2-4 m_j^2\right)B_0(m_H^2,m_j^2,m_j^2) \,,
\end{align}
which appears for all diagrams.
To make explicit the cancellation of UV divergences in the
renormalisation procedure, the UV divergent contributions are extracted
from the integrals according to 
\begin{align}
I^j_{8} &=  \frac{1}{\epsilon} \left(3 m_j^2 - \frac{m_H^2}{2} \right)  + \hat{A}_0(m_j^2)-\frac{1}{2}\left(m_H^2-4 m_j^2\right)\hat{B}_0(m_H^2,m_j^2,m_j^2) \,, \\
	& \equiv \frac{1}{\epsilon} \left(3 m_j^2 - \frac{m_H^2}{2} \right)  + \hat{I}^j_{8} \, .
\end{align}
Therefore, the bare one-loop $h\to b\bar b$ matrix element can be 
written as
\begin{align}
C^{L,(1),\rm bare}_{8,b} &= \frac{1}{v_T}\frac{1}{\epsilon}\Bigg[
4m_b  \left( 3 m_b^2-\frac{m_H^2}{2} \right) \left(C_{qb}^{(1)}+c_{F,3} C_{qb}^{(8)}\right) 
+2 m_{\tau} \left( 3 m_\tau^2-\frac{m_H^2}{2} \right) C^*_{l\tau bq} 
\nonumber \\ &
-m_t \left( 3 m_t^2-\frac{m_H^2}{2} \right) \left( (1+2 N_c) C^{(1)*}_{qtqb} + c_{F,3} C^{(8)*}_{qtqb}\right) \Bigg] +C_{8,b}^{L,(1),\rm fin} \,,
\\[1mm]
C_{8,b}^{L,(1),\rm fin} &= \frac{1}{v_T} \Bigg[ m_b \left(
4 \hat{I}_8^b - 6 m_b^2 + m_H^2\right)\left(C_{qb}^{(1)}+c_{F,3} C_{qb}^{(8)}\right) 
+2 m_\tau \hat{I}_8^{\tau} C^*_{l\tau bq}
\nonumber \\ &
-m_t  \hat{I}_8^t \left( (2 N_c+1) C^{(1)*}_{qtqb} + c_{F,3} C^{(8)*}_{qtqb}\right)  \Bigg] \, .
\end{align}
The corresponding result for $h\to \tautau$ is
\begin{align}
C^{L,(1),\rm bare}_{8,\tau} &=  \frac{1}{v_T}\frac{1}{\epsilon}\Bigg[
4m_{\tau}  \left( 3 m_{\tau}^2-\frac{m_H^2}{2} \right) C_{le}
-2 N_c m_{b} \left( 3 m_b^2-\frac{m_H^2}{2} \right) C^*_{l\tau bq} 
\nonumber \\ &
+2 N_c m_t \left( 3 m_t^2-\frac{m_H^2}{2} \right) C^{(1)*}_{l\tau qt} \Bigg] +C_{8,\tau}^{L,(1),\rm fin} \,,
\\[1mm] 
C_{8,\tau}^{L,(1),\rm fin} &= \frac{1}{v_T} \Bigg[ m_{\tau} \left(4 \hat{I}_{8}^{\tau} - 6 m_{\tau}^2 + m_H^2 \right) C_{le} -2 N_c m_b\hat{I}_{8}^b C^*_{l\tau b q} + 2 N_c m_t\hat{I}_{8}^t C^{(1)*}_{l\tau qt} \Bigg] \, .
\end{align}

\subsection{Counterterms}
\label{sec:4qCT}
As outlined in Section~\ref{sec:OL}, to cancel the poles in the bare
matrix element we must construct the UV counterterms according to 
(\ref{eq:dim6CT}). The four-fermion operators contribute to
operator renormalisation, as well as to fermion mass and wavefunction  
renormalisation.  

The four-fermion contribution to $\delta C^*_{fH}$ is calculated
according to (\ref{eq:delCdef}), where explicit results for
$\dot{C}^*_{fH}$ can be taken from~\cite{Jenkins:2013zja,
  Jenkins:2013wua}.  To adapt those results to the broken phase of the
theory, the Yukawa couplings and the parameter $\lambda$ from the
Higgs potential must be replaced with the physical parameters $m_H$
and $m_f$, as in ~(\ref{eq:Hmass}) and~(\ref{eq:Yuk}) respectively.
Extracting the pieces involving only four-fermion contributions to
$\delta C_{bH}$ and $\delta C_{\tau H}$ gives
\begin{align} \label{delCbH4f}
\delta C^{(4f)}_{bH} &= \frac{\sqrt{2}}{v_T^3} \frac{1}{\epsilon}\bigg[
\frac{1}{2} m_t ( m_H^2 - 4 m_t^2) \left(  (2N_c+1) C^{(1)}_{qtqb} + c_{F,3} C^{(8)}_{qtqb} \right) 
\nonumber \\ &
- 2 m_b ( m_H^2 - 4 m_b^2)  \left(C_{qb}^{(1)} + c_{F,3} C_{qb}^{(8)} \right) 
+ m_{\tau} ( m_H^2 - 4 m_{\tau}^2) C^*_{l\tau bq} \bigg] \,,
\\[1mm]
\delta C^{(4f)}_{\tau H} &= \frac{\sqrt{2}}{v_T^3} \frac{1}{\epsilon}\bigg[
N_c  m_b ( m_H^2 - 4 m_b^2 ) C_{l\tau bq} 
\nonumber \\ &
-2 m_{\tau} ( m_H^2 - 4 m_{\tau}^2) C_{l\tau} - N_c  m_t ( m_H^2 - 4 m_t^2) C^{(1)}_{l\tau qt}
\bigg] \,.
\end{align}

The counterterms from mass and wavefunction renormalisation are
calculated from two-point functions according to~(\ref{eq:delZ})
and~(\ref{eq:delm}).  The relevant one-loop diagrams are of the form
of that shown on the right-hand side of Figure~\ref{fig:4q}.  
The results for the wavefunction and mass
renormalisation for the $b$-quark are
\begin{align}
\label{eq:4fctsB}
 \delta m_b^{(6)} &= \frac{1}{\epsilon}\bigg[  \frac{m_t^3}{2} \left((2 N_c+1) \left(C^{(1)}_{qtqb}+C^{(1)*}_{qtqb}\right)+ c_{F,3} \left(C^{(8)}_{qtqb}+C^{(8)*}_{qtqb}\right) \right) 
 \nonumber \\ &
 - 4 m_b^3\left( C_{qb}^{(1)} + c_{F,3} C_{qb}^{(8)} \right) 
+ m_{\tau}^3 \left(C_{l\tau bq}+C^*_{l\tau bq}\right)  \bigg]  +\delta m_b^{\rm fin}(\mu) \, ,
\nonumber \\[1mm]
 \delta m_b^{\rm fin}(\mu) &=
 \frac{m_t}{2} \hat{A}_0  (m_t^2) \left( (2 N_c+1) \left(C^{(1)}_{qtqb}+C^{(1)*}_{qtqb}\right) + c_{F,3} \left(C^{(8)}_{qtqb}+C^{(8)*}_{qtqb}\right)\right) 
 \nonumber \\ &
+ 2 m_b \left( m_b^2 - 2 \hat{A}_0 (m_b^2) \right) \left( C_{qb}^{(1)} + c_{F,3} C_{qb}^{(8)} \right)
+m_{\tau} \hat{A}_0 (m_{\tau}^2) \left(C_{l\tau bq}+C^*_{l\tau bq}\right) \,,
\nonumber \\[1mm]
\delta Z_b^{(6),L} &=  \frac{1}{\epsilon}\bigg[ 
- \frac{m_t^3}{m_b}  \left( (2 N_c+1) \left(C^{(1)}_{qtqb}-C^{(1)*}_{qtqb}\right) + c_{F,3} \left(C^{(8)}_{qtqb}-C^{(8)*}_{qtqb}\right) \right) 
\nonumber \\ &
+ 2 \frac{m_{\tau}^3}{m_b} \left(C_{l\tau bq}-C^*_{l\tau bq}\right) \bigg]
+ \delta Z_b^{L, \rm fin}(\mu) \,,
\nonumber \\[1mm]
\delta Z_b^{L, {\rm fin}}(\mu) &= 
- \frac{m_t}{m_b} \hat{A}_0 (m_t^2)  \left( (2 N_c+1) \left(C^{(1)}_{qtqb}-C^{(1)*}_{qtqb}\right) + c_{F,3} \left(C^{(8)}_{qtqb}-C^{(8)*}_{qtqb}\right) \right) 
\nonumber \\ &
+ 2 \frac{m_{\tau}}{m_b} \hat{A}_0 (m_{\tau}^2) \left(C_{l\tau bq}-C^*_{l\tau bq}\right) \,,
\nonumber \\[1mm]
\delta Z_b^{(6),R} &= 0 \, .
\end{align}
while those from  $\tau$-leptons are
\begin{align}
\label{eq:4fctsTau}
\delta m^{(6)}_{\tau} &= \frac{1}{\epsilon}\left[ - 4 m_{\tau}^3 C_{l\tau}
+ N_c m_b^3 \left(C_{l\tau bq} + C^*_{l\tau bq} \right)- N_c m_t^3  \left( C^{(1)}_{l\tau qt} + C^{(1)*}_{l\tau qt} \right) \right]
 +\delta m_{\tau}^{\rm fin}(\mu) \,,
 \nonumber \\[1mm]
\delta m_{\tau}^{\rm fin}(\mu) &=
2 m_{\tau} \left( m_{\tau}^2 - 2 \hat{A}_0(m_{\tau}^2) \right) C_{l\tau}
+ N_c m_b \hat{A}_0(m_b^2) \left(C_{l\tau bq}+C^*_{l\tau bq}\right)
\nonumber \\ &
- N_c m_t^3 \hat{A}_0(m_t^2) \left( C^{(1)}_{l\tau qt}+C^{(1)*}_{l\tau qt}\right) \,,
\nonumber \\[1mm]
\delta Z_{\tau}^{(6),L} &=  2 N_c\frac{1}{\epsilon}\bigg[
 \frac{m_t^3}{m_{\tau} }  \left(C^{(1)}_{l\tau qt}-C^{(1)*}_{l\tau qt}\right) 
  - \frac{m_b^3}{m_{\tau} }  \left(C_{l\tau bq}-C^{*}_{l\tau bq}\right) 
\bigg] + \delta Z_{\tau}^{L, \rm fin}(\mu) \,,
\nonumber \\[1mm]
\delta Z_{\tau}^{L, \rm fin}(\mu) &=  2 N_c \bigg(
 \frac{m_t}{m_{\tau} } \hat{A}_0(m_t^2)  \left(C^{(1)}_{l\tau qt}-C^{(1)*}_{l\tau qt}\right)
 -  \frac{m_b}{m_{\tau} }  \hat{A}_0(m_b^2) \left(C_{l\tau bq}-C^{*}_{l\tau bq}\right)
\bigg) \,,
\nonumber \\[1mm]
\delta Z_{\tau}^{(6),R} &= 0 \, . 
\end{align}
Notably, only the real parts of the four-fermion Wilson coefficients contribute to mass
renormalisation, while only the imaginary parts contribute to wavefunction renormalisation.

\subsection{Renormalised matrix element}
\label{sec:4qRes}
Adding together the bare matrix element and UV counterterms as in
(\ref{eq:hbbren}), we find that the UV divergences cancel. We write
the remaining finite contribution as
\begin{align}
i{\cal M}^{(1)}_{8,f}(h\to f\bar f)= 
- i \frac{1}{16\pi^2} \bar u(p_f) 
\left( C_{8,f}^{L,(1)} P_L + C_{8,f}^{R,(1)} P_R\right) v(p_{\bar f})  \, .
\end{align}
The renormalised one-loop matrix element for $h\to b\bar{b}$ decays is
\begin{align} \label{HbbRen4q}
v_T C_{8,b}^{L,(1)} & = 
m_b ( m_H^2 - 4 m_b^2 ) \left( 1 - 2 \hat{B}_0(m_H^2,m_b^2,m_b^2) \right) \left( C_{qb}^{(1)} + c_{F,3} C_{qb}^{(8)} \right)
\nonumber \\ & 
+ m_{\tau} ( m_H^2 - 4 m_{\tau}^2 ) \hat{B}_0(m_H^2,m_{\tau}^2,m_{\tau}^2) C_{l\tau bq}
\nonumber \\ & 
+ \frac{m_t}{2} ( m_H^2 - 4 m_t^2 ) \hat{B}_0(m_H^2,m_t^2,m_t^2)  \left(  (2 N_c +1) C^{(1)*}_{qtqb} + c_{F,3} C^{(8)*}_{qtqb} \right) \,.
\end{align}
The $\mu$-dependence of the one-loop matrix element is contained implicitly in the functions $\hat{B}_0$.  
We can make it explicit by writing
\begin{align}
\hat{B}_0(m_H^2,m_t^2,m_t^2) = \hat{b}_0(m_H^2,m_t^2,m_t^2) - \ln \left( \frac{m_H^2}{\mu^2}\right) \,.
\end{align}
We then find 
\begin{align} \label{HbbRen4qfin}
v_T C_{8,b}^{L,(1)} & = 
m_b ( m_H^2 - 4 m_b^2 ) \left( 1 - 2 \hat{b}_0(m_H^2,m_b^2,m_b^2) \right) \left( C_{qb}^{(1)} + c_{F,3} C_{qb}^{(8)} \right)
\nonumber \\ & 
+ m_{\tau} ( m_H^2 - 4 m_{\tau}^2 ) \hat{b}_0(m_H^2,m_{\tau}^2,m_{\tau}^2) C_{l\tau bq}
\nonumber \\ & 
+ \frac{m_t}{2} ( m_H^2 - 4 m_t^2 ) \hat{b}_0(m_H^2,m_t^2,m_t^2)  \left(  (2 N_c +1) C^{(1)*}_{qtqb} + c_{F,3} C^{(8)*}_{qtqb} \right) 
\nonumber \\ &
-\frac{1}{2}  \frac{v_T^2}{\sqrt{2}} \dot{C}_{bH}^{(4f)*} \ln \left( \frac{m_H^2}{\mu^2} \right) \,.
\end{align}
In obtaining this expression, we have used $\dot{C}_i(\mu) = 2 \epsilon\,\delta C_i(\mu)$
to express~(\ref{delCbH4f}) in a convenient form.
The corresponding result for $h\to \tautau$ decays reads
\begin{align} \label{HtautauRen4q}
v_T C_{8,\tau}^{L,(1)} & = 
m_{\tau} (m_H^2 - 4 m_{\tau}^2)  \left( 1 - 2 \hat{b}_0(m_H^2,m_{\tau}^2,m_{\tau}^2) \right) C_{l\tau}
\nonumber \\ &
+  N_c m_b ( m_H^2 - 4 m_b^2 ) \hat{b}_0(m_H^2,m_{\tau}^2,m_{\tau}^2) C^*_{l\tau bq}
\nonumber \\ &
- N_c m_t ( m_H^2 - 4 m_t^2 ) \hat{b}_0(m_H^2,m_t^2,m_t^2) C^{(1)*}_{l\tau qt}
\nonumber \\ &
-\frac{1}{2}  \frac{v_T^2}{\sqrt{2}} \dot{C}_{\tau H}^{(4f)*} \ln \left( \frac{m_H^2}{\mu^2} \right) \,.
\end{align}
Written in this way, it is clear that the $\mu$-dependence in the
one-loop results arises from the fact that the Wilson coefficients are
renormalised in the \msbar~scheme, and that this $\mu$-dependence
cancels that in the tree-level result (\ref{eq:Mtree}), so that the
renormalised matrix element is $\mu$-independent up to one-loop order.

As the expressions for the scalar integrals appearing
in~(\ref{HbbRen4qfin}) and~(\ref{HtautauRen4q}) are particularly simple, we provide them
explicitly for convenience.  For the contributions from internal
$b$-quark lines, the integral
\begin{align}
\hat{b}_0(m_H^2,m_b^2,m_b^2) = 2 - \bar{z}\, \left[ 
\ln\left(\frac{1+\bar{z}}{1-\bar{z}}\right)- i \pi \right] - \ln\left(\frac{m_b^2}{m_H^2} \right) \, ,
\end{align}
where $\bar{z} = \sqrt{1 - 4 m_b^2/m_H^2}$. The result for internal
$\tau$-lepton lines is then obtained after obvious replacements.  In the case
of top-quark contributions,
\begin{align}
\hat{b}_0(m_H^2,m_t^2,m_t^2) = 2 - 2 z\, {\rm{ArcCot}}\left(z\right) - \ln\left(\frac{m_t^2}{m_H^2} \right) \,,
\end{align}
where $z = \sqrt{4 m_t^2/m_H^2 -1}$.  The right-handed contributions
$C_{8,f}^{R,(1)}$ are obtained as in (\ref{eq:CRs}).  As discussed in
Section~\ref{sec:Inputs}, the quantity $v_T$ in the one-loop results
should be replaced in favour of $G_F$ using (\ref{eq:GF}).  In
fact, neglecting terms of order $\mathcal{O}(1/\Lambda_{\rm NP}^4)$
and higher, the relation $\sqrt{2} v_T^2 = 1/G_F$ can be used.

These results are valid in the on-shell scheme for fermion masses, and
we will use them to study the size of one-loop corrections from
four-fermion operators in Section~\ref{sec:Pheno}. In a more detailed
phenomenological analysis the \msbar~scheme for quark masses may
be preferable. This is particularly true for the $b$-quark, for which
accurate numerical extractions of the $\overline{m}_b(\overline{m}_b)$
exist~\cite{Agashe:2014kda}. At one-loop order, the \msbar~mass is
related to the pole mass according to
\begin{align}
\label{eq:mfbar}
\overline{m}_f(\mu) = m_f + \delta m_f^{\rm fin}(\mu)\,,
\end{align}
where the one-loop results $\delta m_{b,\tau}^{\rm fin}(\mu)$ were given in (\ref{eq:4fctsB})
and  (\ref{eq:4fctsTau}). It is straightforward to obtain the \msbar~results for the 
$h\to f\bar{f}$ matrix element. One eliminates the pole mass $m_f$ in favour of its \msbar~counterpart
using (\ref{eq:mfbar}), and then re-expands the formula as appropriate at one-loop.  One then finds
\begin{align} \label{eq:MSbar}
v_T \overline{C}_{8,f}^{L,(1)} & =  v_T C_{8,f}^{L,(1)}  - \delta m_f^{\rm fin}(\mu)\, .
\end{align}

\section{The one-loop contributions in the large-$m_t$ limit}
\label{sec:MT}
We have obtained the full set of corrections to both $h\to b\bar{b}$
and $h\to \tautau$ decay rates in the limit of vanishing gauge
couplings, and will present them in future work along with the results
which include the full gauge coupling dependence~\cite{longpaper}.  In
this section we focus instead on the leading corrections in the
$m_t\to \infty$ limit, which are a well-defined subset of the full
corrections and potentially dominant numerically.  However, a
number of the interesting features of the renormalisation procedure
are subleading in this limit.  In order to illustrate them, we keep
exact mass dependence of contributions multiplying $1/\epsilon$ poles
and $\mu$-dependent logarithms. In fact, these $\mu$-dependent terms
can be deduced from results for the RG equations of the dimension-6
Wilson coefficients appearing in the tree-level result
(\ref{eq:Mtree}).  These RG equations were calculated explicitly in
the unbroken phase of the theory in \cite{Jenkins:2013zja,
  Jenkins:2013wua}, and our calculation directly in the broken phase
of the theory thus provides a very non-trivial consistency check on
those results, as well as on our renormalisation procedure and
explicit loop calculations.

As with the calculation of four-fermion contributions, we consider
only the third generation contributions.  We additionally make the
assumption of real Wilson coefficients.  To ease the calculation of the
contributing diagrams, of which there are many even with the above
mentioned simplifications, we have implemented the dimension-6
Lagrangian in \texttt{FeynRules}~\cite{Alloul:2013bka}, and
subsequently generated and computed the relevant Feynman diagrams with
\texttt{FeynArts}~\cite{Hahn:2000kx} and
\texttt{FormCalc}~\cite{Hahn:1998yk}.  We give some details of our
procedure for calculating the one-loop corrections mentioned above in
Appendix~\ref{App:LMT}, paying special attention to deriving results
valid in the $m_t\to \infty$ limit.  The renormalised one-loop results
are obtained by evaluating (\ref{eq:hbbren}).  We first give results
for the ingredients entering the counterterms, and then give
the final results for the renormalised one-loop matrix elements at 
the end of the section.

Following the procedure taken for the four-fermion contributions,
we construct the UV counterterms from operator renormalisation by
adapting the results of~\cite{Jenkins:2013zja, Jenkins:2013wua} to
the broken phase of theory. The results are:
\begin{align}\label{delCbH}
\frac{v_T^2}{\sqrt{2}} \delta C_{bH}  &= \frac{1}{\epsilon}\frac{1}{v_T} \bigg[ 
-m_b \left(6 m_b^2+m_H^2\right) \frac{C_{H,\text{kin}}}{v_T^2}
+\frac{1}{4} m_b  \left(2 m_b^2-m_H^2\right) C_{HD}
\nonumber \\ &
+\frac{v_T}{2 \sqrt{2}} \big((10 N_c+21) m_b^2+12 m_H^2+ (6 N_c -3 ) m_t^2 + 6 m_{\tau}^2\big)  C_{bH}
\nonumber \\ &
-\frac{(3-2 N_c) v_T m_b m_t C_{tH}}{\sqrt{2}}
+\sqrt{2} v_T m_b m_{\tau} C_{\tau H} 
- 4 m_b m_{\tau}^2 C_{H\tau}^{(3)}
\nonumber \\ &
- m_b \left(4 N_c m_b^2 - 3 m_H^2+(4 N_c-6) m_t^2\right) C_{Hq}^{(3)} 
+ m_b \left(2 m_b^2+m_H^2\right) \left( C_{Hq}^{(1)} - C_{Hb} \right)
\nonumber \\ &
- m_t \left(-4 N_c m_b^2+m_H^2+2 m_t^2\right) C_{Htb}
\bigg]
+\frac{v_T^2}{\sqrt{2}} \delta C^{(4f)}_{bH} \, ,
\\[1mm] \label{delCeH}
\frac{v_T^2}{\sqrt{2}} \delta C_{\tau H} & = \frac{1}{\epsilon} \frac{1}{v_T} \bigg[
-m_{\tau} \left(6 m_{\tau}^2+m_H^2\right) \frac{C_{H,\text{kin}}}{v_T^2}
+\frac{1}{4} m_{\tau} \left(m_H^2-2 m_{\tau}^2\right) C_{HD}  
\nonumber \\ &
+\frac{v_T}{2\sqrt{2}} \left( 12 m_H^2 + 31 m_{\tau}^2 + 6 N_c \left(m_b^2 + m_t^2\right) \right) C_{\tau H}
+ \sqrt{2} N_c v_T m_{\tau}  ( m_b C_{bH}+ m_t C_{tH} ) 
\nonumber \\ &
+m_{\tau} \bigg( (3 m_H^2 - 4 m_{\tau}^2 ) C^{(3)}_{H\tau} + (m_H^2 + 2 m_{\tau}^2) \left(C^{(1)}_{H\tau}-C_{H\tau}\right) 
\nonumber \\ &
+4 N_c  \left( m_b m_t C_{Htb}  -  (m_b^2 + m_t^2) C^{(3)}_{Hq} \right)
\bigg)  \bigg] 
+ \frac{v_T^2}{\sqrt{2}}\delta C_{\tau H}^{(4f)} \, ,
\\[1mm]
v_T^2\delta C_{HD}  &=\frac{1}{\epsilon}\bigg[
(3 m_H^2 + 4N_c(m_b^2+m_t^2)+4m_\tau^2) C_{HD} + 8 N_c m_b^2 C_{Hb} -8 N_c m_t^2 C_{Ht}
\nonumber \\ &
-8 N_c(m_b^2 - m_t^2) C_{Hq}^{(1)}-8 N_c m_b m_t C_{Htb} 
+ 8m_\tau^2(C_{H\tau}-C_{H\tau}^{(1)})
\bigg] \, ,
\\[1mm] \label{delCHkin}
 \delta C_{H,\text{kin}} &= \left( \delta C_{H\Box} - \frac{\delta C_{HD}}{4} \right) v_T^2 \,,
 \nonumber \\ &
= \frac{1}{\epsilon} \bigg[ 2 \left( 3 m_H^2 + 2 \left( m_{\tau}^2 + N_c \left(m_b^2+m_t^2\right) \right) \right)  \frac{C_{H,\text{kin}}}{v_T^2} + \frac{3}{4} m_H^2 C_{HD}
\nonumber \\ &
-6 \left(m_{\tau}^2 C_{H\tau}^{(3)}  + N_c \left(  (m_b^2 + m_t^2 ) C_{Hq}^{(3)} - m_b m_t C_{Htb} \right) \right) \bigg]\, ,
\\[1mm]
v_T^2\delta C_{HWB}  &= \frac{1}{\epsilon} \left( m_H^2 + 2 \left( m_{\tau}^2 + N_c ( m_b^2 + m_t^2) \right) \right) C_{HWB} \,.
\end{align}

We calculate counterterms from wavefunction, mass, and electric charge
renormalisation from two-point functions as described in
Section~\ref{sec:OL}.  The results for the counterterms (but not for
the renormalised amplitude) are in general gauge-dependent.  We quote
here the results in in 't Hooft-Feynman gauge.\footnote{We have also
  performed the calculation of the renormalised one-loop amplitude in
  unitary gauge and found full agreement with the Feynman gauge
  results.}  They read
\begin{align}
\label{eq:CTs}
\frac{\delta m_b^{(4)}}{m_b} & =  
\frac{C_\epsilon}{v_T^2}\left[ \frac{3}{2\epsilon}\left( m_b^2- m_t^2 \right)-\frac{5}{4}m_t^2\right] \,,
\nonumber \\[1mm]
\delta m_b^{(6)} & =
C_\epsilon\bigg[\frac{1}{\epsilon}\bigg(
3 m_b^3 \frac{C_{H,\text{kin}}}{v_T^2}+\frac{1}{4} m_b^3 C_{HD}
-\frac{3 v_T}{2 \sqrt{2}}\left(2 m_b^2+m_H^2\right) C_{bH}
+m_b^3 \left( C_{Hb} - C_{Hq}^{(1)}\right)
\nonumber \\ &
-3 m_b m_t^2 C_{Hq}^{(3)}
+ m_t^3 C_{Htb} 
-4 m_b^3 \left( C_{qb}^{(1)} + c_{F,3} C_{qb}^{(8)} \right)
+ m_t^3 \left( (2 N_c+1) C_{qtqb}^{(1)}  + c_{F,3} C_{qtqb}^{(8)} \right)
\nonumber \\[1mm]
&+ 2 m_{\tau}^3 C_{l\tau bq} \bigg)
-\frac{5}{2} m_b m_t^2 C_{Hq}^{(3)}+ m_t^3 \left( C_{Htb} + (2 N_c+1) C_{qtqb}^{(1)} +  c_{F,3} C_{qtqb}^{(8)} \right) \bigg]  \,,
\nonumber \\
\frac{\delta m_{\tau}^{(4)}}{m_{\tau}} & =  \frac{C_\epsilon}{v_T^2} \frac{3 m_\tau^2}{2\epsilon}  \,,
\nonumber \\[1mm]
\delta m_{\tau}^{(6)} & = C_\epsilon\bigg[\frac{1}{\epsilon}\bigg( 
3 m_{\tau}^3 \frac{C_{H,{\rm kin}}}{v_T^2} + \frac{1}{4} m_{\tau}^3 C_{HD}
- \frac{3 v_T}{2\sqrt{2}}  (2 m_{\tau}^2 + m_H^2) C_{\tau H} 
+ m_{\tau}^3 \left( C_{H\tau} - C^{(1)}_{H\tau} \right)
\nonumber \\ &
- 4 m_{\tau}^3 C_{l\tau} + 2 N_c \left( m_b^3 C_{l\tau bq} - m_t^3 C^{(1)}_{l\tau qt} \right) \bigg) 
- 2 N_c m_t^3 C^{(1)}_{l\tau qt} \bigg] \,,
\nonumber \\[1mm]
\delta Z_b^{(4),L} & =  \frac{C_\epsilon}{v_T^2}\left[
\frac{1}{\epsilon}\left( -m_b^2- m_t^2 \right)-\frac{3}{2}m_t^2\right] \,,
\nonumber \\[1mm]
\delta Z_b^{(6),L} & =
C_\epsilon\bigg[\frac{1}{\epsilon}\bigg(
- m_b^2 \frac{C_{H, {\rm kin}} }{v_T^2} +\frac{1}{4} m_b^2 C_{HD}
+ \frac{v_T}{\sqrt{2}} m_b C_{bH}
+m_b^2 C_{Hb}+2 \left(m_t^2-m_b^2\right) C_{Hq}^{(3)}
\nonumber \\ &
+ m_b m_t C_{Htb}
\bigg) + m_t^2 C_{Hq}^{(3)} \bigg] \,,
 \nonumber \\[1mm]
\delta Z_{\tau}^{(4),L} & =  -\frac{C_\epsilon}{v_T^2}
\frac{m_{\tau}^2}{\epsilon} \,,
\nonumber \\[1mm]
 \delta Z_{\tau}^{(6),L} & =  \frac{C_\epsilon}{\epsilon}\left[
- m_{\tau}^2 \frac{C_{H,\text{kin}}}{v_T^2}+ \frac{1}{4} m_{\tau}^2 C_{HD} 
+\frac{v_T}{\sqrt{2}} m_{\tau} C_{\tau H} 
+ m_{\tau}^2 \left( C_{H\tau}  - 2  C^{(3)}_{H\tau} \right) \right] \,,
\nonumber \\[1mm]
 \delta Z_b^{(4),R} & = -\frac{C_{\epsilon}}{v_T^2}\frac{2 m_b^2}{\epsilon} \,,
\nonumber \\[1mm]
\delta Z_b^{(6),R} & = \frac{C_\epsilon}{\epsilon}\left[
-m_b^2  \frac{C_{H,\text{kin}}}{v_T^2} +\frac{1}{4} m_b^2 C_{HD}
+\frac{v_T}{\sqrt{2}} m_b C_{bH}
 -m_b^2 \left( C_{Hq}^{(1)}+3 C_{Hq}^{(3)}\right) \right] \,,
\nonumber \\[1mm]
\delta Z_{\tau}^{(4),R} & = -\frac{C_{\epsilon}}{v_T^2}\frac{2 m_{\tau}^2}{\epsilon} \,,
\nonumber \\[1mm]
\delta Z_{\tau}^{(6),R} & = \frac{C_\epsilon}{\epsilon}\left[
- m_{\tau}^2 \frac{C_{H,\text{kin}}}{v_T^2}  + \frac{1}{4} m_{\tau}^2 C_{HD} 
+ \frac{v_T}{\sqrt{2}} m_{\tau} C_{\tau H}  
 - m_{\tau}^2 \left( C^{(1)}_{H\tau}  + 3 C^{(3)}_{H\tau}  \right) \right] \,,
\nonumber \\[1mm]
 \delta Z_h^{(4)}& = \frac{C_\epsilon}{v_T^2}\left[
-\frac{2}{\epsilon}\left( N_c (m_b^2+ m_t^2) + m_{\tau}^2 \right)+\frac{4}{3} N_c m_t^2\right] \,,
\nonumber \\[1mm]
 \delta Z_h^{(6)}& = C_\epsilon\bigg[
\frac{1}{\epsilon}
\bigg( 
- \left( 4 m_{\tau}^2 + 4 N_c \left(m_b^2+m_t^2\right) +14 m_H^2 \right) \frac{C_{H,\text{kin}}}{v_T^2}  -  m_H^2 C_{HD}
\nonumber \\ &
+2 \sqrt{2} v_T \left( N_c \left( m_b C_{bH}+m_t C_{tH}  \right) +  m_{\tau} C_{\tau H} \right)  
\bigg)
+\frac{8}{3} N_c m_t^2 \frac{C_{H,\text{kin}}}{v_T^2}\bigg]  \,,
\nonumber \\[1mm]
\frac{\delta M_W^{(4)}}{M_W}& = 
\frac{C_\epsilon}{v_T^2}\left[
- \frac{1}{\epsilon}\left( N_c ( m_b^2 + m_t^2) + m_{\tau}^2 \right)-\frac{1}{2}N_c m_t^2\right] \,,
\nonumber \\[1mm]
\frac{\delta M_W^{(6)}}{M_W}& =
C_\epsilon\bigg[
\frac{2}{\epsilon}\left( N_c  m_b m_t C_{Htb} - N_c \left(m_b^2+m_t^2\right) C_{Hq}^{(3)} -  m_{\tau}^2 C_{H\tau}^{(3)} \right) - N_c m_t^2 C_{Hq}^{(3)}\bigg] \,,
\nonumber \\[1mm]
\frac{\delta M_Z^{(4)}}{M_Z}& = 
 -\frac{C_\epsilon}{v_T^2}
\frac{1}{\epsilon} \left(  N_c \left(m_b^2 + m_t^2 \right) + m_{\tau}^2 \right) \,,
\nonumber \\[1mm]
\frac{\delta M_Z^{(6)}}{M_Z}& =
 \frac{C_\epsilon}{\epsilon}\bigg[
\frac{1}{4} \left(2 N_c \left(m_b^2+m_t^2\right)+2m_{\tau}^2+3 m_H^2\right) C_{HD}
\nonumber \\ &
+ m_H^2 \hat{c}_w \hat{s}_w C_{HWB}
+ 2 m_{\tau}^2 \left( C_{H\tau} - C_{H\tau}^{(1)} - C_{H\tau}^{(3)} \right)
\nonumber \\ &
+2 N_c \left(  m_b^2 C_{Hb} - m_t^2 C_{Ht} - (m_b^2 - m_t^2) C_{Hq}^{(1)} - (m_b^2 + m_t^2) C_{Hq}^{(3)} \right)
\bigg] \,,
\nonumber \\[1mm]
\frac{\delta\bar{e}^{(4)}}{\bar{e}}&= 0 \,,
\nonumber \\[1mm]
\frac{\delta\bar{e}^{(6)}}{\bar{e}}&= -\frac{C_{\epsilon}}{\epsilon} m_H^2 \hat{c}_w \hat{s}_w C_{HWB} \,.
\end{align} 
We note that the SM results agree with those quoted in \cite{Butenschoen:2007hz}.
Using the results above along with the definition (\ref{eq:dsbar}), we find
\begin{align}
\frac{\delta\bar{s}^{(4)}_w}{\bar{s}_w}&=\frac{\hat{c}_w^2}{2\hat{s}_w^2}\frac{N_c m_t^2}{v_T^2}\,,
\nonumber \\[1mm]
\frac{\delta\bar{s}^{(6)}_w}{\bar{s}_w}& =
\frac{C_{\epsilon}}{\epsilon} \frac{m_H^2 \hat{c}_w}{2 \hat{s}_w} \left( \hat{c}_w^2 - \hat{s}_w^2\right) C_{HWB}
+ \frac{\hat{c}_w^2}{2\hat{s}_w^2} N_c m_t^2 \left[C_{HD}+2 C_{Hq}^{(3)} \right]
\nonumber \\ &
+ \frac{N_c m_t^2 \hat{c}_w}{\hat{s}_w} \left(1- \frac{1}{4 \hat{s}_w^2} \right) C_{HWB}
- \frac{\hat{c}_w \hat{v}_T^2}{4\hat{s}_w} \left(\dot{C}_{HWB} + \frac{\hat{c}_w}{2 \hat{s}_w} \dot{C}_{HD}\right) \ln\left(\frac{m_t^2}{\mu^2}\right) \,.
\end{align}
In the above equations we have defined the quantity
\begin{align}
C_\epsilon = \left(\frac{\mu^2}{m_t^2}\right)^\epsilon =1 
- \epsilon \ln\left(\frac{m_t^2}{\mu^2}\right) + {\cal O}(\epsilon^2)\,.
\end{align}
This is a natural definition for the large-$m_t$ contributions, where only 
logarithms of the form $\ln (m_t^2/\mu^2)$ can appear. When expanding the 
counterterms in $\epsilon$ this definition generates additional, finite
terms which are subleading in the large-$m_t$ limit, i.e. terms of 
the form $m_b^2 \ln m_t^2/\mu^2$.  We choose to keep these subleading terms
in our final analytic results for the renormalised amplitudes, since then
the $\mu$-independence up to one-loop order is manifest, also away from
the large-$m_t$ limit. While these subleading terms may appear more 
naturally as logarithms of the form, e.g. $m_b^2 \ln m_b^2/\mu^2$, the 
argument of the $\mu$-dependent logarithms is not fixed to leading
order in the large-$m_t$ limit considered here --- such ambiguities 
will be resolved by the full calculation \cite{longpaper}.  

\begin{figure}[t]
\centering
\includegraphics[scale=1.2]{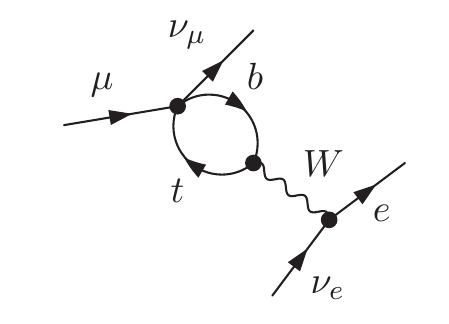}
\hspace{1cm}
\includegraphics[scale=1.2]{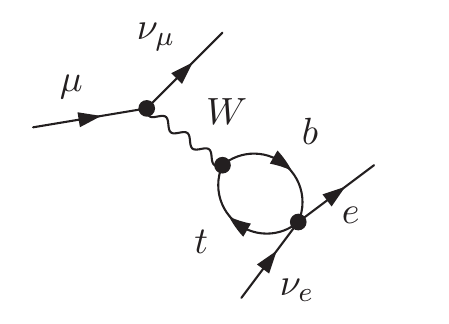}
\caption{Feynman diagrams which contribute to the finite corrections to $\Delta R^{(6,1)}$ in the large-$m_t$ limit.}
\label{fig:GF}
\end{figure}

The full UV counterterm contribution to the
amplitude~(\ref{eq:dim6CT}) can be constructed from these results, and
when added to the bare one-loop matrix element (computed from the sum
of diagrams depicted in Fig.~\ref{fig:Hbb}, for example), one finds
that all UV poles and gauge dependence cancels. The final step in the
calculation is to eliminate $v_T$ dependence using
(\ref{eq:GF_one_loop}).  At one-loop order, we can immediately use the
results above to calculate the expansion coefficients
(\ref{eq:Rcoeffs}) of $\Delta r$.  We find that
\begin{align}
\Delta r^{(4,1)}& = - \frac{\hat{c}_w^2}{\hat{s}_w^2} \frac{N_c m_t^2}{v_T^2} \, , 
\nonumber  \\[1mm]
\Delta r^{(6,1)}& = - \frac{\hat{c}_w^2}{\hat{s}_w^2} N_c m_t^2\left(C_{HD}+2 C_{Hq}^{(3)}\right)
-\frac{N_c m_t^2 \hat{c}_w}{\hat{s}_w}\left(4-\frac{1}{\hat{s}_w^2}\right)C_{HWB}
\nonumber \\
&
+\frac{\hat{c}_w \hat{v}_T^2}{\hat{s}_w}\left(\dot{C}_{HWB}+\frac{\hat{c}_w}{4\hat{s}_w} \dot{C}_{HD}\right)\ln \left(\frac{m_t^2}{\mu^2}\right)\,. 
\end{align}
Obtaining the one-loop contribution $\Delta R^{(6,1)}$ requires a separate but straightforward calculation.  In
the large-$m_t$ limit, the only non-logarithmic finite contributions arise from the insertions of four-fermion
operators onto the usual tree-level $W$ boson exchange graph, resulting
in the diagrams in Figure~\ref{fig:GF}.
%
Evaluating those diagrams in the large-$m_t$ limit (and including the
$\mu$-dependent terms implied by the RG equations)
we find that 
\begin{align} \Delta R^{(6,1)} = \frac{N_c m_t^2}{\sqrt{2} v_T^2}
 \left( C^{(3)}_{\substack{lq \\ \mu\mu33}} +
  C^{(3)}_{\substack{lq \\ ee33}} \right)
   -\frac{1}{2\sqrt{2}}
\left(\dot{C}_{\substack{Hl \\ ee}}^{(3)}+\dot{C}_{\substack{Hl \\
      \mu\mu}}^{(3)} - \frac{1}{2}\left(\dot{C}_{\substack{ll \\ \mu
        ee \mu}} + \dot{C}_{\substack{ll \\ e\mu \mu e}} \right)
\right) \ln \left( \frac{m_t^2}{\mu^2} \right) \, . 
\end{align}

We are now in position to give the final results for the renormalised
one-loop corrections in the large-$m_t$ limit.  We write the results
in terms of expansion coefficients as
\begin{align}
\label{eq:OneLoopCoeffs}
i{\cal M}(h\to f\bar f)= 
-i\bar u(p_f)  v(p_{\bar f})\left[ A_f^{(4,0)}+ A_f^{(6,0)}
+\frac{1}{16\pi^2}\left(A_f^{(4,1)} 
+ A_f^{(6,1)}\right) \right] \,.
\end{align}
The results for $h\to b\bar b$ are, at tree-level, 
\begin{align} \label{H2bb1tree}
A^{(4,0)}_b &= \left(\sqrt{2} G_F\right)^{\frac{1}{2}} m_b \,,
\\[1mm]
A^{(6,0)}_b &= A^{(4,0)}_b C_{H,{\rm kin}} -  \frac{C_{bH}}{2 G_F} 
+ A^{(4,0)}_b \frac{ \Delta R^{(6,0)}}{2G_F} \,.
\end{align}
The one-loop results are
\begin{align} \label{H2bb1loop}
A^{(4,1)}_b &= A_b^{(4,0)} G_F  m_t^2 \left( \frac{-18+7 N_c}{3\sqrt{2}}\right) \,,
\\[1mm]
A^{(6,1)}_b &= 
A_b^{(4,0)} m_t^2 \left( 
\frac{3 G_F}{\sqrt{2}} (-2+N_c) C_{H,\text{kin}}
+ \left(-1+N_c \right) C_{Hq}^{(3)}
\right)
+\frac{(-15+4 N_c) m_t^2 }{12}  \frac{C_{bH}}{\sqrt{2}}
\nonumber \\ &
+ \frac{1}{2}\left[A_b^{(4,0)} \dot{C}_{H,\text{kin}} - 
\frac{1}{2 G_F}\dot{C}_{bH}\right]\ln\left(\frac{m_t^2}{\mu^2}\right)  \nonumber \\ 
& + 
\frac{1}{2G_F} \left(A^{(4,0)}_b  \Delta R^{(6,1)} + 3 A^{(4,1)}_b \Delta R^{(6,0)}\right)
\,. \label{eq:A61b}
\end{align}
Similarly, for $h\to \tautau$, at tree-level, 
\begin{align}
A^{(4,0)}_{\tau} &= \left(\sqrt{2} G_F\right)^{\frac{1}{2}} m_{\tau} \,,
\\[1mm]
A^{(6,0)}_{\tau} &= A^{(4,0)}_{\tau} C_{H,{\rm kin}} - \frac{C_{\tau H}}{2 G_F} 
+ A^{(4,0)}_{\tau} \frac{ \Delta R^{(6,0)}}{2G_F}
\,. 
\end{align}
The one-loop results are
\begin{align}
A^{(4,1)}_{\tau} &= A_{\tau}^{(4,0)} G_F m_t^2 \frac{7 N_c}{3\sqrt{2}} \,,
\\[1mm]
A^{(6,1)}_{\tau} &= 
A_{\tau}^{(4,0)} N_c m_t^2 \left(  \frac{3}{\sqrt{2}} G_F C_{H,\text{kin}} + C_{Hq}^{(3)} \right)
+\frac{N_c m_t^2}{3}\frac{C_{\tau H}}{\sqrt{2}}
 \nonumber \\ &
+ \frac{1}{2}\left[A_{\tau}^{(4,0)} \dot{C}_{H,\text{kin}} - 
\frac{1}{2 G_F}\dot{C}_{\tau H}\right]\ln\left(\frac{m_t^2}{\mu^2}\right) \nonumber \\ 
& + 
\frac{1}{2G_F} \left(A^{(4,0)}_{\tau}  \Delta R^{(6,1)} + 3 A^{(4,1)}_{\tau} \Delta R^{(6,0)}\right)
\label{eq:A61tau}
\,.
\end{align}

The main results of this section are the renormalised one-loop
contributions to the decay amplitudes in (\ref{eq:A61b})
and~(\ref{eq:A61tau}). We have written them in a form which makes
clear that the renormalised decay amplitudes are $\mu$-independent up
to one-loop.  We have calculated the coefficients of the
$\mu$-dependent logarithms directly in the broken phase of the theory,
and emphasise that it is a non-trivial cross-check that are results are consistent
with those by the RG equations of \cite{Jenkins:2013zja,
  Jenkins:2013wua}.  The non-logarithmic one-loop corrections, on the other
hand, cannot be deduced from RG equations and are a new result.
Interestingly, the potentially dominant non-logarithmic contributions
proportional to $m_t^3$ occurring in the bare matrix elements are
cancelled by those in the mass renormalisation counterterms 
$\delta m_{b,\tau}^{(6)}$ in (\ref{eq:CTs}).  Obviously, this is a 
scheme-dependent result that would not hold if these masses were instead
renormalised in the \msbar~scheme.

\section{Impact on phenomenology}
\label{sec:Pheno}

In this section we explore the implications of the one-loop
corrections provided in~(\ref{eq:A61b}) and~(\ref{eq:A61tau}) on the
interpretation of Higgs decay data.  
We begin by commenting on the application of RG-improved perturbation
theory to the interpretation of data from experiment, before discussing the
sensitivity of Higgs decay measurements to various Wilson coefficients. 

In the context of the current calculation, the relevant physical scale for the process is $\mu =\mu_t \sim m_t$.
By setting the scale which appears in both one-loop and tree-level dimension-six amplitudes
to this value, the large logarithms which appear in the one-loop matrix 
elements in~(\ref{eq:A61b}) and~(\ref{eq:A61tau}) are absorbed into the Wilson coefficients $C_i(\mu_t)$.
The decay rates which are then computed from squaring
the sum of these amplitudes are then a function of purely 
finite terms and Wilson coefficients defined at the scale $C_i(\mu_t)$.
Constraints on the possible values of these Wilson coefficients 
(defined at the scale $\mu_t$) can then be obtained by 
performing a fit to the available data.\footnote{In the situation
where a global fit is performed to a large data set, which may involve differential 
measurements or processes with different scales, a relevant scale 
should be chosen for each data point and the constraints obtained
on the Wilson coefficients in this way should be presented at a common scale.}

In the (hopeful) scenario where such a fit prefers non-zero values for some of
these Wilson coefficients $C_i(\mu_t)$, it will be possible to interpret such a scenario in 
terms of new physics. This can be done without making reference to a specific model, by evolving 
these Wilson coefficients from the scale $\mu_t$ to the scale $\Lambda_{\rm NP}$ by solving
the RG equations. In doing so, potentially large logarithms are resummed into RG evolution
factors which relate Wilson coefficients at different scales.
In fact, provided the scale $\Lambda_{\rm NP}$ does not exceed the  
scale $\mu_t$ by several orders of magnitude, the relation
between Wilson coefficients at different scales can be approximated 
through the one-loop solution to the RG equations:
\begin{align} 
\label{RG}
C_i(\mu_t) = C_i(\Lambda_{\rm NP}) +\frac{1}{2}\frac{1}{16\pi^2} 
\dot{C}_i(\Lambda_{\rm NP}) \ln \left( \frac{\mu_t^2}{\Lambda_{\rm NP}^2} \right) \,.
\end{align}
The constraints on the values of the Wilson coefficients $C_i(\mu_t)$ obtained
in this way are therefore translated into constraints on the Wilson coefficients
defined at the scale $\Lambda_{\rm NP}$. 
The benefit of such an approach is that solution to~(\ref{RG}) is fully known
at one-loop~\cite{Jenkins:2013zja, Jenkins:2013wua,  Alonso:2013hga}. 
Therefore, it is possible to directly test specific new physics models
by matching them to the SMEFT at a scale $\mu \sim \Lambda_{\rm NP}$, and
comparing the consistency of the set of non-vanishing Wilson coefficients 
$C_i(\Lambda_{\rm NP})$ generated in this matching procedure with those obtained 
from data (having been evolved to the scale $\Lambda_{\rm NP}$).
In general, the main goal of NLO calculations within the SMEFT is to evaluate
the purely finite contributions to the one-loop amplitudes, as we have 
provided in~(\ref{eq:A61b}) and~(\ref{eq:A61tau}).
The importance of evaluating these contributions is to determine
whether or not they have any impact on the extraction of the values
of the $C_i(\mu_t)$.
Of course the NLO calculations also provide a  
cross check of the previous anomalous dimension calculations.

In the following, we will assess the impact our results on the decay 
rates at the scale $\mu_t$, and then briefly discuss the potential 
interpretation of a non-zero extraction of Wilson coefficients at this 
scale in terms of new physics in a model independent way.
We calculate decay rates as a double expansion in loop factors and 
$1/\Lambda_{\rm NP}$, neglecting self-interference of dimension-6 operators
as well as that of one-loop contributions.  We thus decompose 
the decay rate as   
\begin{align}
\Gamma(h\to f\bar{f}) = B_f\left[\Gamma_f^{(4,0)} + \Gamma_f^{(6,0)}
+ \Gamma_f^{(4,1)} + \Gamma_f^{(6,1)}\right]+\dots \, ,
\end{align}  
where 
\begin{align}
B_{\tau} = \frac{m_H}{8 \pi} \left(1-\frac{4 m_\tau^2}{m_H^2}\right)^{\frac{3}{2}} \, , \qquad B_b = N_c  \frac{m_H}{8 \pi} \left(1-\frac{4 m_b^2}{m_H^2}\right)^{\frac{3}{2}} \, .
\end{align}
Taking advantage of the fact that the matrix elements are real,
we can define the SM contributions by
\begin{align}
\Gamma_f^{(4,0)} &=  \left[ A_f^{(4,0)} \cdot A_f^{(4,0)} \right] \, , \qquad
\Gamma_f^{(4,1)} =  \frac{1}{16\pi^2} \left[ 2 A_f^{(4,0)} \cdot A_f^{(4,1)} \right] \, ,
\end{align}
while the dimension-6 contributions are 
\begin{align}
\Gamma_f^{(6,0)} & =   \left[ 2 A_f^{(4,0)} \cdot A_f^{(6,0)} \right] \, , \qquad
\Gamma_f^{(6,1)} = \frac{1}{16\pi^2} \left[ 2 \left(A_f^{(6,0)} \cdot A_f^{(4,1)}+ A_f^{(4,0)} \cdot A_f^{(6,1)}\right)  \right] \, .
\end{align}
To evaluate these expression numerically, we use the following set of
input parameters: $m_t = 173.3$~GeV, $m_b = 4.75$~GeV, $m_{\tau} =
1.777$~GeV, $m_H = 125.0$~GeV, $G_{F} =
1.16638\cdot10^{-5}$~GeV$^{-2}$.  
To make the suppression of dimension-6 contributions explicit, 
we define the dimensionless quantities
\begin{align} \label{WCdef}
C_i(\mu_t) \equiv \frac{\tilde{C}_i(\mu_t)}{\Lambda_{\rm NP}^2} \, , \qquad 
C_i(\Lambda_{\rm NP}) \equiv \frac{\hat{C}_i(\Lambda_{\rm NP})}{\Lambda_{\rm NP}^2} \,.
\end{align}
%

\subsection{$h\to b\bar b$ decays}
We now consider the relative size of the different types of corrections
for the process $h\rightarrow b \bar{b}$.  First of all, the ratio
of tree-level dimension-6 and SM contributions is given by
\begin{align} \label{LOrat0}
\frac{\Gamma_b^{(6,0)}}{\Gamma_b^{(4,0)}} = -  \frac{1}{G_F (\sqrt{2} G_F)^{\frac{1}{2}} m_b}\frac{\tilde{C}_{bH}}{\Lambda_{\rm NP}^2}  + \frac{1}{G_F} \left( \sqrt{2} \left(\frac{\tilde{C}_{H\Box}}{\Lambda_{\rm NP}^2}-\frac{1}{4}\frac{\tilde{C}_{HD}}{\Lambda_{\rm NP}^2}\right) + \frac{\Delta \tilde{R}^{(6,0)}}{\Lambda_{\rm NP}^2} \right) \,,
\end{align}
where the explicit definition of $C_{H,{\rm kin}}$
in~(\ref{eq:Class3}) has been used.  Numerically, at a scale of
$\Lambda_{\rm NP} = 1$~TeV, this amounts to
\begin{align} \label{LOrat}
 \frac{\Gamma_b^{(6,0)}}{\Gamma_b^{(4,0)}} = -
4.44 \tilde{C}_{bH} + 0.03 \left( 4 \tilde{C}_{H\Box} - \tilde{C}_{HD}\right)
+ 0.09 \Delta \tilde{R}^{(6,0)}
\,.  
\end{align}
The dimension-6 contributions are large if $\tilde{C}_{bH} \sim {\cal O}(1)$.  
On the other hand, if one assumes that $\tilde{C}_{bH}\sim
y_b$, as would be the case in MFV, then the result is 
\begin{align} \label{LOratb}
 \frac{\Gamma_b^{(6,0)}}{\Gamma_b^{(4,0)}} = 
 -0.12 \frac{\tilde{C}_{bH}}{y_b} + ... \,, 
\end{align}
where the ellipses denote the remaining terms of~(\ref{LOrat}), 
whose sensitivity is numerically comparable in an MFV-like scenario.

We next study the size of one-loop corrections.  The ratio of the one-loop
to tree-level corrections in the SM is
\begin{align}
\frac{\Gamma_b^{(4,1)}}{\Gamma_b^{(4,0)}} = \frac{G_F m_t^2 }{8\pi^2} 
\left( \frac{-18+7 N_c}{3\sqrt{2}}\right) = 0.003 \,,
\end{align}
in agreement with previous
results~\cite{Kniehl:1991ze}.
The one-loop corrections are quite small due to a large cancellation
between the $N_c$-dependent and $N_c$-independent terms in
the numerator.

The ratio of the one-loop SMEFT and tree-level SM predictions
can also be obtained in a similar fashion, written in terms of Wilson 
coefficients defined at the scale $\mu_t$, we find
\begin{align} \label{eq:loopratNLO}
\nonumber
\frac{\Gamma_b^{(6,1)}}{\Gamma_b^{(4,0)}} &= 
- \frac{m_t^2}{(\sqrt{2} G_F)^{\frac{1}{2}} m_b}  \frac{(-21+10 N_c)}{96\pi^2\sqrt{2}} \frac{\tilde{C}_{bH}}{\Lambda_{\rm NP}^2}
-m_t^2 \frac{(9-4 N_c)}{48\pi^2} \left( 4 \frac{\tilde{C}_{H\Box}}{\Lambda_{\rm NP}^2}- \frac{\tilde{C}_{HD}}{\Lambda_{\rm NP}^2} \right)
\\ \nonumber 
& \hspace{0.5cm}+m_t^2 \frac{(-1+N_c)}{8 \pi^2} \frac{\tilde{C}^{(3)}_{Hq}}{\Lambda_{\rm NP}^2}  + \frac{1}{16\pi^2 G_F}\frac{\Delta \tilde{R}^{(6,1)}}{\Lambda_{\rm NP}^2} + m_t^2\frac{(-18+7 N_c)}{12\pi^2\sqrt{2}}\frac{\Delta \tilde{R}^{(6,0)}}{\Lambda_{\rm NP}^2} \,,
\\[1mm] \nonumber & 
\simeq -0.01 \tilde{C}_{bH} + 10^{-3} \left( 0.19 \left(4 \tilde{C}_{H\Box} - \tilde{C}_{HD} + 4 \tilde{C}^{(3)}_{Hq}\right) + 0.54\left(\Delta \tilde{R}^{(6,0)}+\Delta \tilde{R}^{(6,1)}  \right) \right) \, ,
\\[1mm] &
\simeq -0.0003 \frac{\tilde{C}_{bH}}{y_b}+... \,.
\end{align}
By comparing the numerical pre-factor of $\tilde{C}_{bH}$ in the expression above
with that in the corresponding LO expression~(\ref{LOrat}), we see that these
finite one-loop corrections are numerically unimportant. 
As noted in the previous section, the potentially dominant
non-logarithmic terms proportional to $m_t^3$ in the bare
matrix element and multiplying the $C_{Htb}$, $C^{(1)}_{qtqb}$ and
$C^{(8)}_{qtqb}$ coefficients are cancelled exactly by the mass
counterterm for the $b$-quark in the on-shell scheme.
Our calculation therefore justifies using a LO SMEFT analysis,
at least in the on-shell scheme,
to constrain the Wilson coefficients $\tilde{C}_i/\Lambda_{\rm NP}^2$
appearing in~(\ref{LOrat}), and then in turn using an RG analysis
to interpret such constraints at the scale $\Lambda_{\rm NP}$.
In fact, the anomalous dimension calculation which is required
to do such an analysis has already been presented~\cite{Elias-Miro:2013mua}, 
where the authors also studied the phenomenological implications 
of their results.

We perform a similar analysis below by expressing the Higgs decay
rate in terms of Wilson coefficients at the scale $\Lambda_{\rm NP} \sim 1$~TeV,
where we use the hatted notation for the Wilson coefficients $\hat{C}_i(\Lambda_{\rm NP})$
introduced in~(\ref{WCdef}) to differentiate them from $\tilde{C}_i(\mu_t)$.
We therefore compute a compact expression for the ratio of the SMEFT
decay rate with respect to the tree-level SM prediction. Retaining only the 
numerically important terms, which correspond to those which appear at tree-level 
and in addition a subset of the mixing contributions generated by the running of $C_{bH}$, we find
%
\begin{align}
\label{eq:loopratbb}
\frac{\Gamma_b^{(6,0)}+\Gamma_b^{(6,1)}}{\Gamma_b^{(4,0)}} &\simeq  
\nonumber
-  \frac{1}{G_F (\sqrt{2} G_F)^{\frac{1}{2}} m_b}\frac{\hat{C}_{bH}}{\Lambda_{\rm NP}^2}  + \frac{1}{G_F} \left( \sqrt{2} \left(\frac{\hat{C}_{H\Box}}{\Lambda_{\rm NP}^2}-\frac{1}{4}\frac{\hat{C}_{HD}}{\Lambda_{\rm NP}^2}\right) + \frac{\Delta \hat{R}^{(6,0)}}{\Lambda_{\rm NP}^2} \right)
\\  &
+
\frac{1}{16 \pi^2} \frac{1}{\Lambda_{\rm NP}^2}\Bigg[
 \frac{3}{\sqrt{2}(\sqrt{2} G_F)^{\frac{1}{2}}m_b } \left( 4 m_H^2 + m_t^2 (-1 + 2N_c) \right) \hat{C}_{bH}  \nonumber \\ &
 \hspace{2.5cm} - 2 \frac{m_t}{m_b} (m_H^2 + 2 m_t^2 ) \hat{C}_{Htb}  -\frac{m_t}{m_b} (4 m_t^2 - m_H^2) 
 \nonumber \\ &
\hspace{2.5cm} \left( (2 N_c+1) \hat{C}^{(1)}_{qtqb} + c_{F,3} \hat{C}^{(8)}_{qtqb} \right) \bigg)
 \Bigg] \ln\left(\frac{\Lambda_{\rm NP}^2}{m_t^2}\right) \,.
\end{align}
Evaluating (\ref{eq:loopratbb}) at the
scale $\Lambda_{\rm NP} = 1$~TeV, we find
\begin{align} \label{UVscale}
\nonumber
\frac{\Gamma_b^{(6,0)}+\Gamma_b^{(6,1)}}{\Gamma_b^{(4,0)}} \simeq  &-3.93 \hat{C}_{bH} 
- 0.59 \hat{C}^{(1)}_{qtqb}
- 0.12 \hat{C}_{Htb}
+0.12\hat{C}_{H\Box}
- 0.11 \hat{C}^{(8)}_{qtqb}
\\ &
+ 0.09 \Delta \hat{R}^{(6,0)}
- 0.03\hat{C}_{HD}
\, .
\end{align}

The above analyses demonstrates that this decay rate is numerically most 
sensitive to the coefficients $\hat{C}_{bH}$ and $\hat{C}^{(1)}_{qtqb}$. 
%
Interestingly, these particular Wilson coefficients are also not well experimentally constrained.
For instance, the four-fermion operator multiplying $C^{(1)}_{qtqb}$ 
does not contribute to $Zbb$ couplings at one-loop level due to its Dirac structure.
This can be observed by direct calculation, or by examining the anomalous
dimension matrix of the Wilson coefficients $C^{(1)}_{Hb}$ and $C^{(3)}_{Hq}$
which alter the $Z$ boson couplings to fermions~\cite{Jenkins:2013wua}. 
Consequently, $C^{(1)}_{qtqb}$ is not subject to strong constraints from LEP data.
Nor does the operator $Q^{(1)}_{qtqb}$ give large contributions to top-quark pair 
production at hadron colliders, since the tree-level partonic process
$b\bar{b}\to t\bar{t}$ is highly suppressed as result of the exceedingly 
small $b\bar{b}$-quark PDF luminosity.
This leads us to consider a simplified analysis, where all Wilson
coefficients except for $C_{bH}$ and $C^{(1)}_{qtqb}$, which are
currently unconstrained phenomenologically, vanish.  

\begin{figure}[t]
\centering
\includegraphics[scale=0.7]{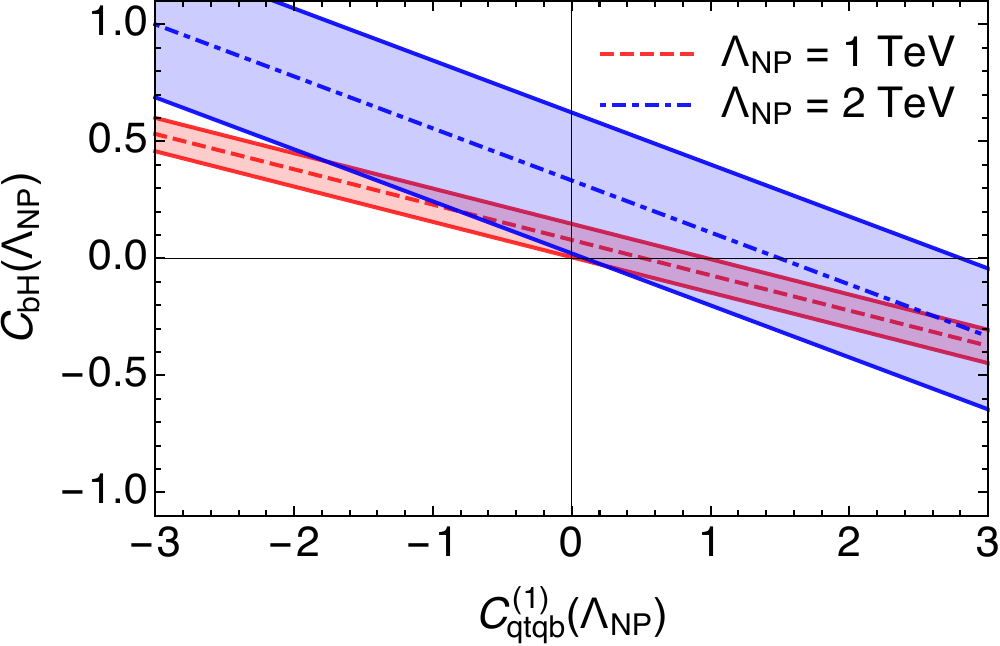} \hspace{0.5cm}
\includegraphics[scale=0.7]{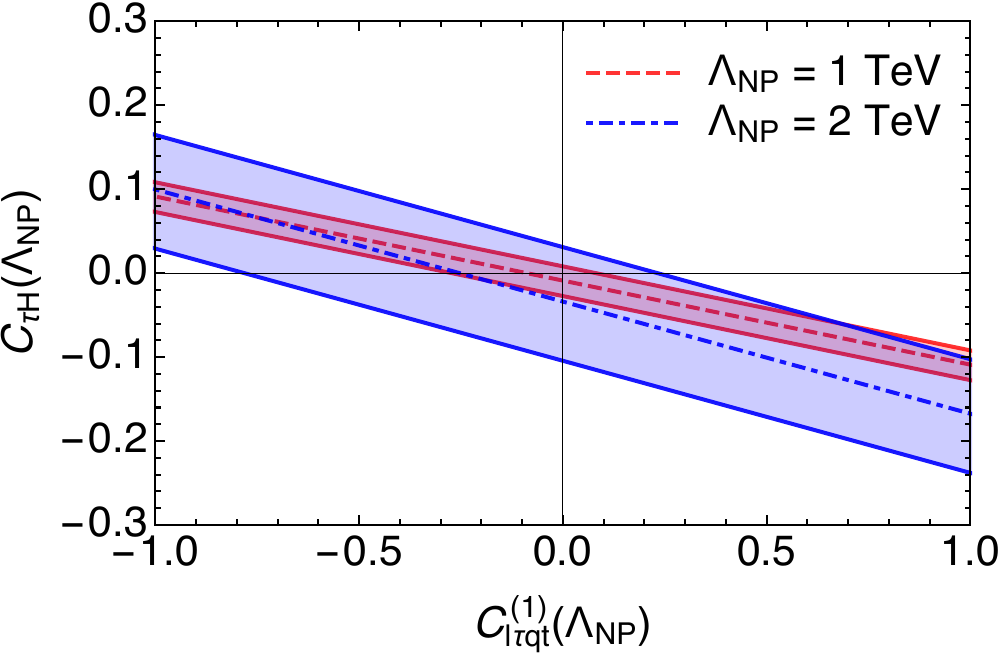}
\caption{\label{fig:CbH} Constraints in the plane of the Wilson coefficients $C_{bH}(\Lambda_{\rm NP})-C^{(1)}_{qtqb}(\Lambda_{\rm NP})$ (left) and $C_{\tau H}(\Lambda_{\rm NP})-C^{(1)}_{l\tau qt}(\Lambda_{\rm NP})$ (right), based on a simplified analysis of the combined CMS and ATLAS Run-I data~\cite{CMS-PAS-HIG-15-002}.}
\end{figure}

Under such conditions, it also straightforward to place experimental
constraints on these Wilson coefficients by including the available 
Higgs decay data~\cite{CMS-PAS-HIG-15-002}. To do so, we can
identify the extracted signal strength $\mu^{bb}$ with the SMEFT and SM
decay rates in the following way $\mu^{bb} = 1+\Gamma_b^{(6)}/\Gamma_b^{(4)}$.
Under the assumption that new physics does not alter Higgs boson production,
we can use the experimental extraction of $\mu^{bb}$ from the combined CMS and ATLAS
analysis of $\mu^{bb} = 0.69^{+0.29}_{-0.27}$. Using the tree-level formula~(\ref{LOrat0}),
leads to the following constraint
\beq
\frac{\tilde{C}_{bH}(\mu_t)}{G_F \Lambda_{\rm NP}^2} = \left(5.98_{-5.59}^{+5.20}\right)\times 10^{-3} \,.
\eeq
To interpret the impact of the measurement of $\mu^{bb}$ in terms of Wilson coefficients defined 
at the scale $\Lambda_{\rm NP}$, we can simply use the compact formula~(\ref{eq:loopratbb}) assuming 
non-vanishing $C_{bH}$ and $C^{(1)}_{qtqb}$ Wilson coefficients. The solution, which depends 
both linearly and logarithmically on the choice of $\Lambda_{\rm NP}$, is presented for the 
choices $\Lambda_{\rm NP} = 1, 2$~TeV in Figure~\ref{fig:CbH} 
(along side the results for the $h\to\tautau$ which will be discussed below).
Interestingly, the available data already constrains the values of these Wilson coefficients 
to be $\mathcal{O}(1)$, and prefers positive values of both $\hat{C}_{bH}$ and $\hat{C}^{(1)}_{qtqb}$
to accommodate the slightly low value of $\mu^{bb}$ observed in data. It should be noted that
zero values of these Wilson coefficients are consistent with the data at 1$\sigma$~CL.

In the above scenario, we have placed constraints on the Wilson coefficients without
reference to any particular UV completion. However, in a broad range of UV completions, 
such as those studied in~\cite{Elias-Miro:2013mua}, these Wilson coefficients are expected 
to scale as $\hat{C}^{\rm MFV}_{bH}\sim y_b \hat{C}_{bH}$ and $\hat{C}_{qtqb}^{\rm MFV,(1)}\sim y_b y_t \hat{C}_{qtqb}^{(1)}$, 
where the $\hat{C}_i$ are order one quantities (we have so far referred to such a scenario as
MFV-like).
In this case, useful bounds on the rescaled coefficients $\hat{C}_i$ can be 
expected only once experimental measurements improve in precision by at 
least an order of magnitude.\footnote{For further, model-dependent phenomenological 
studies in new physics scenarios such as supersymmetry, we refer the reader to~\cite{Elias-Miro:2013mua}.}
Such precision may be experimentally challenging at the LHC, even with the large amount of
data expected during Run II~\cite{Dawson:2013bba}. However, such precision
can be achieved at an $e^+e^-$ machine~\cite{Baer:2013cma,Fujii:2015jha,Gomez-Ceballos:2013zzn,Fan:2014vta},
where sub-percent level precision is estimated for both $b$-quark and $\tau$-lepton 
final states for particular $e^+e^-$ programs.

\subsection{$h\to \tau\bar{\tau}$ decays}
The analysis for the case of $h\to \tau\bar{\tau}$ decays proceeds along 
similar lines. In fact, the ratio of the tree-level dimension-6 contributions
to the tree-level SM contributions is given by~(\ref{LOratb}) 
after replacing $b\to\tau$.  The one-loop corrections in the SM
are instead
\begin{align}
\frac{\Gamma_\tau^{(4,1)}}{\Gamma_\tau^{(4,0)}} = \frac{G_F m_t^2 }{8\pi^2} 
\left(\frac{7 N_c}{3\sqrt{2}}\right) = 0.022 \,,
\end{align}
Compared to the decay into $b$-quarks, this contribution is generated
solely by the finite terms present in the counterterm for $\delta M_W$
and $\delta Z_{h}$, and so there is no large cancellation.  

To assess the impact of the purely finite one-loop SMEFT corrections, we take
the ratio of this decay rate (at the scale $\mu_t$) with that of the tree-level SM decay rate, finding
\begin{align} \label{eq:loopratNLOtau}
\nonumber
\frac{\Gamma_{\tau}^{(6,1)}}{\Gamma_{\tau}^{(4,0)}} &= 
- \frac{5 N_c m_t^2}{48\pi^2(2 \sqrt{2} G_F)^{\frac{1}{2}} m_{\tau}}   \frac{\tilde{C}_{\tau H}}{\Lambda_{\rm NP}^2}
+ \frac{N_c m_t^2}{12\pi^2} \left( 4 \frac{\tilde{C}_{H\Box}}{\Lambda_{\rm NP}^2}- \frac{\tilde{C}_{HD}}{\Lambda_{\rm NP}^2} \right)
\\ \nonumber 
& \hspace{0.5cm}+ \frac{N_c m_t^2}{8 \pi^2} \frac{\tilde{C}^{(3)}_{Hq}}{\Lambda_{\rm NP}^2}  + \frac{1}{16\pi^2 G_F}\frac{\Delta \tilde{R}^{(6,1)}}{\Lambda_{\rm NP}^2} + \frac{7 N_c m_t^2}{12\pi^2\sqrt{2}}\frac{\Delta \tilde{R}^{(6,0)}}{\Lambda_{\rm NP}^2} \,,
\\[1mm] \nonumber & 
\simeq -0.09 \tilde{C}_{\tau H} + 10^{-2} \left( 0.08 \left(4 \tilde{C}_{H\Box} - \tilde{C}_{HD}\right) + 0.11 \tilde{C}^{(3)}_{Hq} + 0.38 \Delta \tilde{R}^{(6,0)}+0.054\Delta \tilde{R}^{(6,1)}  \right) \, ,
\\[1mm] &
\simeq -0.0009 \frac{\tilde{C}_{\tau H}}{y_{\tau}}+... \,.
\end{align}
In this case, the finite corrections are more important than in the
$h\to b\bar{b}$ decay, particularly for $\tilde{C}_{\tau H}$. However, 
in analogy to the $h\to b\bar{b}$ decay, the potentially 
numerically large term proportional to $m_t^3$ in the bare matrix element (which multiplies
$\tilde{C}^{(1)}_{l\tau qt}$) is exactly cancelled by that in the $\tau$-lepton mass counter-term.
%
%
%
We can also perform a simplified analysis for $\tau$-leptons, under the assumption that
only non-zero values for the Wilson coefficients $\hat{C}_{\tau H}$ and $\hat{C}^{(1)}_{l\tau qt}$
are allowed. In this case, we use the experimentally extracted value of $\mu^{\tau\tau}$ from 
the combined CMS and ATLAS analysis, given by $\mu^{\tau\tau} = 1.12^{+0.25}_{-0.23}$. We find 
\begin{align}
\frac{\tilde{C}_{\tau H}(\mu_t)}{G_F \Lambda_{\rm NP}^2} &=  \left( -0.87^{+1.66}_{-1.80} \right) \times 10^{-3} \qquad {\rm LO}\,, \\[1mm]
\frac{\tilde{C}_{\tau H}(\mu_t)}{G_F \Lambda_{\rm NP}^2} &=  \left( -0.86^{+1.65}_{-1.79} \right) \times 10^{-3} \qquad {\rm LMT}\,.
\end{align}
In the first case (labelled ${\rm LO}$) the Wilson coefficient is extracted using the LO formula $\Gamma_{\tau}^{(i)} = \Gamma_{\tau}^{(i,0)}$. In the latter (labelled ${\rm LMT}$), the large $m_t$-limit one-loop corrections are also included in the extraction as $\Gamma_{\tau}^{(i)} = \Gamma_{\tau}^{(i,0)}+\Gamma_{\tau}^{(i,1)}$.
This demonstrates that the purely finite corrections are indeed not important for the interpretation
of the experimental data in this case either.
To extract constraints on the Wilson coefficients at the scale $\Lambda_{\rm NP}$ directly from $\mu^{\tautau}$, 
we also provide a general compact analytic expression in terms of Wilson-coefficients defined at the scale $\Lambda_{\rm NP}$.
\begin{align} \label{eq:looprattau}
\nonumber 
\frac{\Gamma_{\tau}^{(6,0)}+\Gamma_{\tau}^{(6,1)}}{\Gamma_{\tau}^{(4,0)}} &\simeq  
-  \frac{1}{G_F (\sqrt{2} G_F)^{\frac{1}{2}} m_{\tau}}\frac{\hat{C}_{\tau H}}{\Lambda_{\rm NP}^2}  + \frac{1}{G_F} \left( \sqrt{2} \left(\frac{\hat{C}_{H\Box}}{\Lambda_{\rm NP}^2}-\frac{1}{4}\frac{\hat{C}_{HD}}{\Lambda_{\rm NP}^2}\right) + \frac{\Delta \hat{R}^{(6,0)}}{\Lambda_{\rm NP}^2} \right) \\ \nonumber &
 + \frac{1}{16 \pi^2} \frac{1}{\Lambda_{\rm NP}^2}\Bigg[ -\frac{5 N_c m_t^2}{3\sqrt{2}(\sqrt{2} G_F)^{\frac{1}{2}} m_{\tau}}  \hat{C}_{\tau H}
 -\Bigg( \frac{6 ( 2 m_H^2 + N_c m_t^2) }{\sqrt{2}(\sqrt{2} G_F)^{\frac{1}{2}} m_{\tau}} \hat{C}_{\tau H}
  \nonumber \\ &
 \hspace{2.5cm}+  \frac{m_t}{m_{\tau}}    
	 2 N_c (4 m_t^2 - m_H^2) \hat{C}^{(1)}_{l\tau qt}\Bigg) \ln\left(\frac{\Lambda_{\rm NP}^2}{m_t^2}\right) 
 \Bigg] \, .
 \end{align}
The extracted values of $\hat{C}_{\tau H}(\Lambda_{\rm NP})$ of 
$\hat{C}^{(1)}_{l\tau qt}(\Lambda_{\rm NP})$, assuming otherwise
vanishing Wilson coefficients, are presented in the right plot of Figure~\ref{fig:CbH}.
Once again, we have chosen provide the solutions for the scale choices $\Lambda_{\rm NP} = 1, 2$~TeV.

We therefore arrive at similar conclusions for $\tau$-leptons as in
the case for $b$-quarks. That is, measurements of the Higgs decay 
rate already provide (better than) $\mathcal{O}(1)$ constraints on the combination
of the Wilson coefficients $\hat{C}_{\tau H}(\Lambda_{\rm NP})$ and 
$\hat{C}^{(1)}_{l\tau qt}(\Lambda_{\rm NP})$ for values of $\Lambda_{\rm NP}$ in the
few TeV range.
In a scenario where these Wilson coefficients scale as $\hat{C}^{\rm MFV}_{\tau H}\sim y_{\tau} \hat{C}_{\tau H}$ 
and $\hat{C}_{l\tau qt}^{{\rm MFV}, (1)}\sim y_{\tau} y_t \hat{C}_{l\tau qt}^{(1)}$, measurements
in a clean $e^+e^-$ environment will be necessary to constrain $\mathcal{O}(1)$ values 
of $\hat{C}_{\tau H}$ and $\hat{C}_{l\tau qt}^{(1)}$.

\section{Conclusions}
\label{sec:conclusions}

We have calculated a set of one-loop
corrections to $h\to b\bar{b}$ and $h\to \tau\bar{\tau}$ decay rates
within the SMEFT.  In particular, we gave exact one-loop results from
four-fermion operators, and in addition the leading electroweak
corrections in the large-$m_t$ limit.  We also calculated the one-loop
corrections to muon decay in the same limit, which is necessary to
implement the $G_F$ input-parameter scheme.

Our SMEFT calculations were carried out within an extension of the
on-shell renormalisation scheme for electroweak corrections, which was
described in Section~\ref{sec:OL}.  In this procedure counterterms
related to wavefunction, mass, and electric charge renormalisation are
determined from one-loop two-point functions directly in the broken
phase of the theory as in the on-shell scheme used in SM calculations.
These counterterms receive contributions from both SM and dimension-6
operators, which we calculated explicitly within the approximations
described above.  The counterterms related to operator
renormalisation, on the other hand, are defined in the \msbar~scheme
and constructed using results from RG equations for Wilson
coefficients determined in the unbroken phase of the theory.  While
the idea behind this procedure is simple, the specifics are rather
involved, and we have shown explicitly how it correctly cancels
UV-divergent contributions from a total of 21 different operators in
the limit of vanishing gauge couplings. As a non-trivial check of 
our results, we also computed the coefficients of all $\mu$-dependent
logarithms in this same limit, and showed explicitly that they
have the form dictated by the RG equations.
%

In Section~\ref{sec:Pheno} we assessed the impact of the SM and SMEFT contributions to the $h\to b\bar{b}$ and $h\to \tau\bar{\tau}$ decay rates. To do so, we compute the ratios of the decay rates in SMEFT, for Wilson coefficients defined at the scale $\mu = m_t$, with respect to the SM. We find that potentially large non-logarithmic contributions of $\mathcal{O}(m_t^2)$ are numerically unimportant. In particular, $m_t^3$ contributions in the bare matrix element multiplying poorly constrained four-fermion scalar operators are cancelled exactly by those those appearing in the on-shell mass counterterms. The current analysis suggests, at least in the large-$m_t$ limit, that a simplified
leading-logarithmic analysis is sufficient.  That is, one can calculate the decay rate matrix elements at  LO  in the SMEFT, use the results to constrain Wilson coefficients at the scale $\mu \sim m_t$, and then use one-loop RG 
equations to translate the results into constraints at the scale $\mu \sim \Lambda_{\rm NP}$. 
We emphasise that it is still important to obtain the full NLO expression for the SMEFT decay rate, and such a computation is currently in progress~\cite{Gauld:2016kuu,longpaper}.
In the meantime, we have performed a such simplified analysis focussing on the Wilson coefficients $C_{fH}(\mu_t)$, which are not subject to strong experimental constraints.
Finally, making use of the RG equations, compact analytic formulas for the $b$-quark and $\tau$-lepton decay rates in terms of Wilson
coefficients at the generic scale $\Lambda_{\mathrm{NP}}$ are provided. For values of $\Lambda_{\mathrm{NP}}$ in the several TeV range, these decay rates are mostly sensitive to the pairs of Wilson coefficients $C_{bH}(\Lambda_{\mathrm{NP}})-C^{(1)}_{qtqb}(\Lambda_{\mathrm{NP}})$ (for $b$-quarks) and $C_{\tau H}(\Lambda_{\mathrm{NP}})-C^{(1)}_{l \tau q t}(\Lambda_{\mathrm{NP}})$ (for $\tau$-leptons).
Within the current experimental precision for these decay rates~\cite{CMS-PAS-HIG-15-002}, $\mathcal{O}(1)$ constraints can be placed on these pairs of Wilson coefficients as shown in Fig.~\ref{fig:CbH}. 
We note that if instead these Wilson coefficients scale according to $\hat{C}^{\rm MFV}_{fH}\sim y_f \hat{C}_{fH}$, $\hat{C}_{qtqb}^{\rm MFV,(1)}\sim y_b y_t \hat{C}_{qtqb}^{(1)}$ and $\hat{C}_{l\tau qt}^{\rm MFV,(1)}\sim y_{\tau} y_t \hat{C}_{l\tau qt}^{(1)}$, as may be expected in a scenario with MFV, then the simplified analysis applied in Section~\ref{sec:Pheno} is clearly not adequate.
In the potential future scenario where sub-percent precision is achievable for measurements of
Higgs decay rates~\cite{Baer:2013cma,Fujii:2015jha,Gomez-Ceballos:2013zzn,Fan:2014vta}, 
it will become increasingly important to include information from multiple processes, and in addition to improve
the precision of the calculations which enter a global fit to dimension-6 Wilson coefficients. 
Extending SMEFT calculations to NLO will improve the accuracy
of the theoretical predictions allowing for a more precise comparison of data in terms of 
non-vanishing Wilson coefficients.

\section*{Acknowledgements}
The research of D.~J.~S. is supported by an STFC Postgraduate Studentship.
We are grateful to Ulrich Haisch and Michael Trott for useful discussions.
We also thank Celine Degrande for support with \texttt{FeynRules}.

\appendix

\section{The Large-$m_t$ limit}
\label{App:LMT}

To clarify the procedure of taking the large-$m_t$ limit, we consider
the calculation of the following contribution to the
process $h\rightarrow b\bar b$ which appears in Feynman gauge
\begin{figure}[ht]
\centering
\includegraphics[scale=1.0]{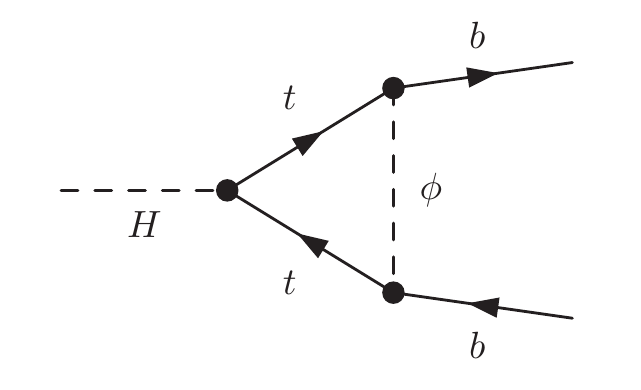}
\label{fig:LMT}
\end{figure}

\noindent We consider both the SM, as well as the contribution due to the
Class 7 operator $Q_{Hq}^{(3)}$, $(H^\dag i\overleftrightarrow{D}^I_\mu H)(\bar q_p \tau^I \gamma^\mu q_r)$. For the diagram considered here this operator alters the coupling of the quarks to the Goldstone bosons.
The amplitude for this diagram is (setting $V_{tb}$ to unity)
\beq
\begin{aligned}
\mathcal{A} = i \frac{2 m_t}{v_T^3} &\int \frac{d^d l}{(2\pi)^d} \bar{u}(p_b)  \Bigg[  
\left[ m_b P_L - m_t P_R  -  v_T^2 ( \slashed{l} - \slashed{p}_b)  C^{(3)}_{Hq} P_L \right] 
\frac{1}{(l-p_b)^2 - m_W^2}
\\ &
\frac{\slashed{l} + m_t}{l^2-m_t^2} \frac{\left(\slashed{l}-\slashed{p}_b-\slashed{p}_{\bar{b}}\right)+m_t}{(l-p_b-p_{\bar{b}})^2 - m_t^2} 
\left[ m_t P_L - m_b P_R + v_T^2 ( \slashed{l} - \slashed{p}_b)  C^{(3)}_{Hq} P_L \right]  \Bigg] v(p_{\bar{b}})
\end{aligned}
\eeq

\noindent After performing the reduction to scalar integrals this can 
be written as 
\beq
\mathcal{A} = - i  \bar{u}(p_b) v(p_{\bar{b}}) \frac{1}{16 \pi^2}\frac{m_b}{v_T^3} \left( \mathcal{A}^{\rm{div.}} +  \mathcal{A}^{\rm{fin.}}\right)
\eeq
The divergent contribution is
\beq
\mathcal{A}^{\rm{div.}} = \frac{m_t^2}{\epsilon}\left( 2 + 5 v_T^2 C^{(3)}_{Hq} \right) \, .
\eeq
\noindent Separating the finite SM and dimension-6 contributions, we find
\begin{align} \label{eq:LMTfull}
\mathcal{A}^{\rm{fin.(4)}} = &  \frac{2 m_t^2}{4 m_b^2 - m_H^2} \Bigg[ 
2 (m_b^2 - m_t^2)  \hat{B}_0(m_b^2,m_t^2,m_W^2)
\nonumber \\ &
+ \left( 2 m_b^2 - m_H^2 + 2 m_t^2 \right) \hat{B}_0(m_H^2,m_t^2,m_t^2)
\nonumber  \\ &
- \left( 2 \left(m_b^2 - m_t^2\right)^2 + m_W^2 \left( m_H^2 - 2 m_b^2 - 2 m_t^2 \right) \right) C_0( m_H^2, m_b^2, m_b^2, m_t^2, m_t^2,m_W^2)\Bigg] \, ,
\nonumber \\
\mathcal{A}^{\rm{fin.(6)}} = & v_T^2 C^{(3)}_{Hq} \Bigg[
\frac{m_t^2}{m_b^2} \left( \hat{A}_0(m_t^2) - \hat{A}_0(m_W^2) \right)
+ \frac{4 m_t^2}{4 m_b^2 -  m_H^2} (2 m_b^2 - m_H^2 + 2 m_t^2 ) \hat{B}_0(m_H^2,m_t^2,m_t^2)
\nonumber \\ &
+\frac{1}{4 m_b^2 -  m_H^2} \bigg( \frac{m_t^2}{m_b^2} \left( m_b^2 \left( 12 (m_b^2-m_t^2) - m_H^2 + 4 m_W^2 \right) +m_H^2 \left(m_t^2 - m_W^2\right) \right) \hat{B}_0(m_b^2,m_t^2,m_W^2)
\nonumber \\ &
-4 m_t^2 \left( 2 \left(m_b^2 - m_t^2\right)^2 + m_W^2 \left( m_H^2 - 2 m_b^2 - 2 m_t^2 \right) \right) C_0( m_H^2, m_b^2, m_b^2, m_t^2, m_t^2,m_W^2) \bigg) \Bigg] \,,
\end{align}
where $\hat{A}_0(s)$ is the finite part of the integral $A_0(s)$ defined in~(\ref{eq:ABints}).
We define the large-$m_t$ limit of the finite parts of the integrals appearing in the results as a series in $1/m_t$. The large-$m_t$ limit of the $\hat{A}_0(m_t^2)$ integral is trivial, 
and the corresponding limit of the $\hat{B}_0$ scalar integrals appearing
in these expressions can be obtained by expanding~(\ref{eq:Bint}) as
\begin{align} \label{eq:Bexp}
\lim_{m_t \rightarrow \infty} \hat{B}_0(m_1^2,m_t^2,m_2^2) &\rightarrow 1 + \frac{1}{m_t^2} \left( \frac{m_1^2}{2} + m_2^2 \ln \left[ \frac{m_2^2}{m_t^2} \right] \right) - \ln \left[ \frac{m_t^2}{\mu^2} \right] \, , \\
\lim_{m_t \rightarrow \infty} \hat{B}_0(m_1^2,m_t^2,m_t^2) &\rightarrow \frac{ m_1^2}{6 m_t^2} - \ln \left[ \frac{m_t^2}{\mu^2} \right] \, .
\end{align}
For the triangle integrals, it is possible to further simplify the integrals appearing in the amplitude by ignoring all fermion masses, except that of the top quark. This is a suitable simplification to make in the context of the current phenomenological study.
In the limit of vanishing gauge couplings, where contributions  proportional to positive powers of $M_W^2$ can be neglected, the evaluation of the $C_0$ functions is further simplified. Note that neither of these limits introduces any extra singularities in the triangle integrals used. Explicitly we have,
\beq
\lim_{m_t \rightarrow \infty} C_0( m_H^2, m_b^2, m_b^2, m_t^2, m_t^2,m_W^2) \rightarrow \lim_{m_t \rightarrow \infty} C_0( m_H^2, 0,0, m_t^2, m_t^2,0) \, ,
\eeq
and similarly for other $C_0$ functions. The integral appearing in~(\ref{eq:LMTfull}) 
becomes
\beq
\lim_{m_t \rightarrow \infty} C_0( m_H^2, 0, 0, m_t^2, m_t^2, 0)  \rightarrow -\frac{1}{m_t^2} - \frac{m_H^2}{12 m_t^4} \, ,
\eeq
while in the full calculation we also make use of
\beq
\lim_{m_t \rightarrow \infty} C_0( 0, m_H^2, 0, m_t^2, 0,0)  \rightarrow -\frac{1}{m_t^2}\left(1 + i \pi + \ln \left[\frac{m_t^2}{m_H^2} \right] \right) \, ,
\eeq
both of which are obtained from~\cite{Kniehl:1991ze}.
Thus, we find the expressions for the finite contributions appearing in~(\ref{eq:LMTfull})
simplify to
\begin{align}
\lim_{m_t \rightarrow \infty} \mathcal{A}^{\rm{fin.(4)}} & \rightarrow -m_t^2 \left( 1  + 2 \ln \left[ \frac{m_t^2}{\mu^2} \right] \right) \, .
\nonumber \\
\lim_{m_t \rightarrow \infty} \mathcal{A}^{\rm{fin.(6)}} & \rightarrow -m_t^2 v_T^2 C^{(3)}_{Hq} \left( \frac{3}{2}  + 5 \ln \left[ \frac{m_t^2}{\mu^2} \right] \right) \, .
\end{align}
The procedure outlined above is applied to all the finite corrections provided
in Section~\ref{sec:MT}.

\newpage
\begin{table}
\begin{center}
\small
\begin{minipage}[t]{4.45cm}
\renewcommand{\arraystretch}{1.5}
\begin{tabular}[t]{c|c}
\multicolumn{2}{c}{$1:X^3$} \\
\hline
$Q_G$                & $f^{ABC} G_\mu^{A\nu} G_\nu^{B\rho} G_\rho^{C\mu} $ \\
$Q_{\widetilde G}$          & $f^{ABC} \widetilde G_\mu^{A\nu} G_\nu^{B\rho} G_\rho^{C\mu} $ \\
$Q_W$                & $\epsilon^{IJK} W_\mu^{I\nu} W_\nu^{J\rho} W_\rho^{K\mu}$ \\ 
$Q_{\widetilde W}$          & $\epsilon^{IJK} \widetilde W_\mu^{I\nu} W_\nu^{J\rho} W_\rho^{K\mu}$ \\
\end{tabular}
\end{minipage}
\begin{minipage}[t]{2.7cm}
\renewcommand{\arraystretch}{1.5}
\begin{tabular}[t]{c|c}
\multicolumn{2}{c}{$2:H^6$} \\
\hline
$Q_H$       & $(H^\dag H)^3$ 
\end{tabular}
\end{minipage}
\begin{minipage}[t]{5.1cm}
\renewcommand{\arraystretch}{1.5}
\begin{tabular}[t]{c|c}
\multicolumn{2}{c}{$3:H^4 D^2$} \\
\hline
$Q_{H\Box}$ & $(H^\dag H)\Box(H^\dag H)$ \\
$Q_{H D}$   & $\ \left(H^\dag D_\mu H\right)^* \left(H^\dag D_\mu H\right)$ 
\end{tabular}
\end{minipage}
\begin{minipage}[t]{2.7cm}

\renewcommand{\arraystretch}{1.5}
\begin{tabular}[t]{c|c}
\multicolumn{2}{c}{$5: \psi^2H^3 + \hbox{h.c.}$} \\
\hline
$Q_{eH}$           & $(H^\dag H)(\bar l_p e_r H)$ \\
$Q_{uH}$          & $(H^\dag H)(\bar q_p u_r \widetilde H )$ \\
$Q_{dH}$           & $(H^\dag H)(\bar q_p d_r H)$\\
\end{tabular}
\end{minipage}

\vspace{0.25cm}

\begin{minipage}[t]{4.7cm}
\renewcommand{\arraystretch}{1.5}
\begin{tabular}[t]{c|c}
\multicolumn{2}{c}{$4:X^2H^2$} \\
\hline
$Q_{H G}$     & $H^\dag H\, G^A_{\mu\nu} G^{A\mu\nu}$ \\
$Q_{H\widetilde G}$         & $H^\dag H\, \widetilde G^A_{\mu\nu} G^{A\mu\nu}$ \\
$Q_{H W}$     & $H^\dag H\, W^I_{\mu\nu} W^{I\mu\nu}$ \\
$Q_{H\widetilde W}$         & $H^\dag H\, \widetilde W^I_{\mu\nu} W^{I\mu\nu}$ \\
$Q_{H B}$     & $ H^\dag H\, B_{\mu\nu} B^{\mu\nu}$ \\
$Q_{H\widetilde B}$         & $H^\dag H\, \widetilde B_{\mu\nu} B^{\mu\nu}$ \\
$Q_{H WB}$     & $ H^\dag \tau^I H\, W^I_{\mu\nu} B^{\mu\nu}$ \\
$Q_{H\widetilde W B}$         & $H^\dag \tau^I H\, \widetilde W^I_{\mu\nu} B^{\mu\nu}$ 
\end{tabular}
\end{minipage}
\begin{minipage}[t]{5.2cm}
\renewcommand{\arraystretch}{1.5}
\begin{tabular}[t]{c|c}
\multicolumn{2}{c}{$6:\psi^2 XH+\hbox{h.c.}$} \\
\hline
$Q_{eW}$      & $(\bar l_p \sigma^{\mu\nu} e_r) \tau^I H W_{\mu\nu}^I$ \\
$Q_{eB}$        & $(\bar l_p \sigma^{\mu\nu} e_r) H B_{\mu\nu}$ \\
$Q_{uG}$        & $(\bar q_p \sigma^{\mu\nu} T^A u_r) \widetilde H \, G_{\mu\nu}^A$ \\
$Q_{uW}$        & $(\bar q_p \sigma^{\mu\nu} u_r) \tau^I \widetilde H \, W_{\mu\nu}^I$ \\
$Q_{uB}$        & $(\bar q_p \sigma^{\mu\nu} u_r) \widetilde H \, B_{\mu\nu}$ \\
$Q_{dG}$        & $(\bar q_p \sigma^{\mu\nu} T^A d_r) H\, G_{\mu\nu}^A$ \\
$Q_{dW}$         & $(\bar q_p \sigma^{\mu\nu} d_r) \tau^I H\, W_{\mu\nu}^I$ \\
$Q_{dB}$        & $(\bar q_p \sigma^{\mu\nu} d_r) H\, B_{\mu\nu}$ 
\end{tabular}
\end{minipage}
\begin{minipage}[t]{5.4cm}
\renewcommand{\arraystretch}{1.5}
\begin{tabular}[t]{c|c}
\multicolumn{2}{c}{$7:\psi^2H^2 D$} \\
\hline
$Q_{H l}^{(1)}$      & $(H^\dag i\overleftrightarrow{D}_\mu H)(\bar l_p \gamma^\mu l_r)$\\
$Q_{H l}^{(3)}$      & $(H^\dag i\overleftrightarrow{D}^I_\mu H)(\bar l_p \tau^I \gamma^\mu l_r)$\\
$Q_{H e}$            & $(H^\dag i\overleftrightarrow{D}_\mu H)(\bar e_p \gamma^\mu e_r)$\\
$Q_{H q}^{(1)}$      & $(H^\dag i\overleftrightarrow{D}_\mu H)(\bar q_p \gamma^\mu q_r)$\\
$Q_{H q}^{(3)}$      & $(H^\dag i\overleftrightarrow{D}^I_\mu H)(\bar q_p \tau^I \gamma^\mu q_r)$\\
$Q_{H u}$            & $(H^\dag i\overleftrightarrow{D}_\mu H)(\bar u_p \gamma^\mu u_r)$\\
$Q_{H d}$            & $(H^\dag i\overleftrightarrow{D}_\mu H)(\bar d_p \gamma^\mu d_r)$\\
$Q_{H u d}$ + h.c.   & $i(\widetilde H ^\dag D_\mu H)(\bar u_p \gamma^\mu d_r)$\\
\end{tabular}
\end{minipage}

\vspace{0.25cm}

\begin{minipage}[t]{4.75cm}
\renewcommand{\arraystretch}{1.5}
\begin{tabular}[t]{c|c}
\multicolumn{2}{c}{$8:(\bar LL)(\bar LL)$} \\
\hline
$Q_{ll}$        & $(\bar l_p \gamma_\mu l_r)(\bar l_s \gamma^\mu l_t)$ \\
$Q_{qq}^{(1)}$  & $(\bar q_p \gamma_\mu q_r)(\bar q_s \gamma^\mu q_t)$ \\
$Q_{qq}^{(3)}$  & $(\bar q_p \gamma_\mu \tau^I q_r)(\bar q_s \gamma^\mu \tau^I q_t)$ \\
$Q_{lq}^{(1)}$                & $(\bar l_p \gamma_\mu l_r)(\bar q_s \gamma^\mu q_t)$ \\
$Q_{lq}^{(3)}$                & $(\bar l_p \gamma_\mu \tau^I l_r)(\bar q_s \gamma^\mu \tau^I q_t)$ 
\end{tabular}
\end{minipage}
\begin{minipage}[t]{5.25cm}
\renewcommand{\arraystretch}{1.5}
\begin{tabular}[t]{c|c}
\multicolumn{2}{c}{$8:(\bar RR)(\bar RR)$} \\
\hline
$Q_{ee}$               & $(\bar e_p \gamma_\mu e_r)(\bar e_s \gamma^\mu e_t)$ \\
$Q_{uu}$        & $(\bar u_p \gamma_\mu u_r)(\bar u_s \gamma^\mu u_t)$ \\
$Q_{dd}$        & $(\bar d_p \gamma_\mu d_r)(\bar d_s \gamma^\mu d_t)$ \\
$Q_{eu}$                      & $(\bar e_p \gamma_\mu e_r)(\bar u_s \gamma^\mu u_t)$ \\
$Q_{ed}$                      & $(\bar e_p \gamma_\mu e_r)(\bar d_s\gamma^\mu d_t)$ \\
$Q_{ud}^{(1)}$                & $(\bar u_p \gamma_\mu u_r)(\bar d_s \gamma^\mu d_t)$ \\
$Q_{ud}^{(8)}$                & $(\bar u_p \gamma_\mu T^A u_r)(\bar d_s \gamma^\mu T^A d_t)$ \\
\end{tabular}
\end{minipage}
\begin{minipage}[t]{4.75cm}
\renewcommand{\arraystretch}{1.5}
\begin{tabular}[t]{c|c}
\multicolumn{2}{c}{$8:(\bar LL)(\bar RR)$} \\
\hline
$Q_{le}$               & $(\bar l_p \gamma_\mu l_r)(\bar e_s \gamma^\mu e_t)$ \\
$Q_{lu}$               & $(\bar l_p \gamma_\mu l_r)(\bar u_s \gamma^\mu u_t)$ \\
$Q_{ld}$               & $(\bar l_p \gamma_\mu l_r)(\bar d_s \gamma^\mu d_t)$ \\
$Q_{qe}$               & $(\bar q_p \gamma_\mu q_r)(\bar e_s \gamma^\mu e_t)$ \\
$Q_{qu}^{(1)}$         & $(\bar q_p \gamma_\mu q_r)(\bar u_s \gamma^\mu u_t)$ \\ 
$Q_{qu}^{(8)}$         & $(\bar q_p \gamma_\mu T^A q_r)(\bar u_s \gamma^\mu T^A u_t)$ \\ 
$Q_{qd}^{(1)}$ & $(\bar q_p \gamma_\mu q_r)(\bar d_s \gamma^\mu d_t)$ \\
$Q_{qd}^{(8)}$ & $(\bar q_p \gamma_\mu T^A q_r)(\bar d_s \gamma^\mu T^A d_t)$\\
\end{tabular}
\end{minipage}

\vspace{0.25cm}

\begin{minipage}[t]{3.75cm}
\renewcommand{\arraystretch}{1.5}
\begin{tabular}[t]{c|c}
\multicolumn{2}{c}{$8:(\bar LR)(\bar RL)+\hbox{h.c.}$} \\
\hline
$Q_{ledq}$ & $(\bar l_p^j e_r)(\bar d_s q_{tj})$ 
\end{tabular}
\end{minipage}
\begin{minipage}[t]{5.5cm}
\renewcommand{\arraystretch}{1.5}
\begin{tabular}[t]{c|c}
\multicolumn{2}{c}{$8:(\bar LR)(\bar L R)+\hbox{h.c.}$} \\
\hline
$Q_{quqd}^{(1)}$ & $(\bar q_p^j u_r) \epsilon_{jk} (\bar q_s^k d_t)$ \\
$Q_{quqd}^{(8)}$ & $(\bar q_p^j T^A u_r) \epsilon_{jk} (\bar q_s^k T^A d_t)$ \\
$Q_{lequ}^{(1)}$ & $(\bar l_p^j e_r) \epsilon_{jk} (\bar q_s^k u_t)$ \\
$Q_{lequ}^{(3)}$ & $(\bar l_p^j \sigma_{\mu\nu} e_r) \epsilon_{jk} (\bar q_s^k \sigma^{\mu\nu} u_t)$
\end{tabular}
\end{minipage}
\end{center}
\caption{\label{op59}
The 59 independent dimension-6 operators built from Standard Model fields which conserve baryon number, as given in Ref.~\cite{Grzadkowski:2010es}. The operators are divided into eight classes: $X^3$, $H^6$, etc. Operators with $+\hbox{h.c.}$ in the table heading also have hermitian conjugates, as does the $\psi^2H^2D$ operator $Q_{Hud}$. The subscripts $p,r,s,t$ are flavour indices, The notation is described in \cite{Jenkins:2013zja}.
}
\end{table}

\begin{figure}[t]
\centering
\includegraphics[scale=1.0]{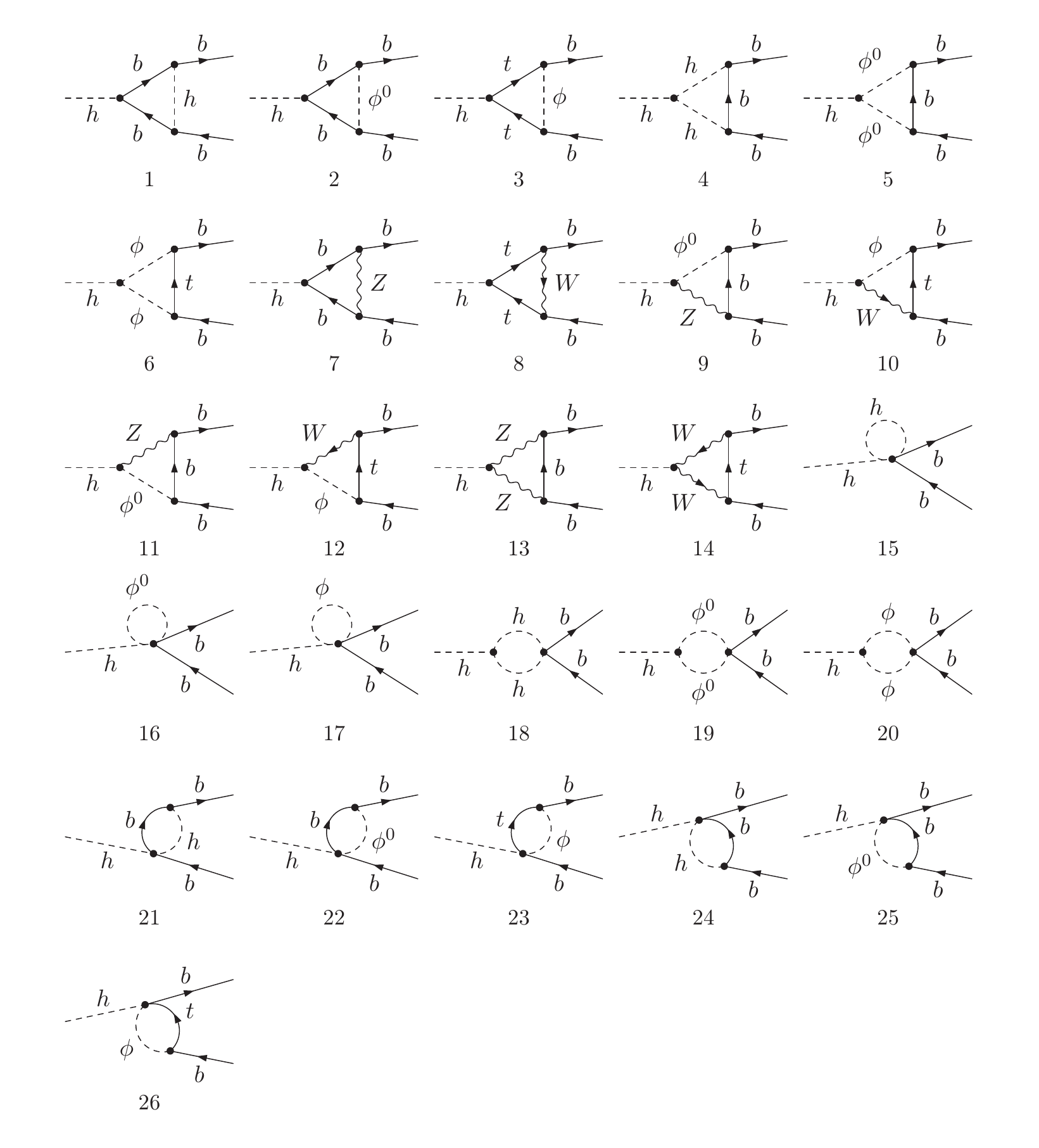}
\caption{Representative diagrams contributing to the process $h\to b\bar b$ in 't Hooft - Feynman gauge. 
The dimension-6 contributions are inserted onto the relevant vertices, and contributions to $\mathcal{O}(1/\Lambda_{NP}^2)$ are kept.
The fields $\phi$, $\phi^0$ refer to the charged and neutral Goldstone bosons. In addition to the usual SM diagrams, note the presence of Diagrams 15-17 which are generated solely by Class 5 operators. The contributions from Class 8 operators are depicted on the left side of Fig.~\ref{fig:4q}.}
\label{fig:Hbb}
\end{figure}

\bibliographystyle{JHEP}
\input{H2ff_LMT.bbl}

\end{document}

%% file: H2ff_LMT.bbl
\providecommand{\href}[2]{#2}\begingroup\raggedright\endgroup